\documentclass[ajl]{emulateapj}

\usepackage{graphicx,epsfig,natbib,color}
\usepackage{lscape}
\usepackage{graphicx}
\usepackage{epsfig}
\usepackage{natbib}
\usepackage[colorlinks, citecolor=blue]{hyperref}

\slugcomment{Accepted for publication in ApJ Supplement Series}

\begin{document}
\title{On the Characteristics of Footpoints of Solar Magnetic Flux Ropes during the Eruption}

\author{X. Cheng$^{1,2}$ \& M. D. Ding$^{1,2}$}

\affil{$^1$School of Astronomy and Space Science, Nanjing University, Nanjing 210093, China}\email{xincheng@nju.edu.cn}
\affil{$^2$Key Laboratory for Modern Astronomy and Astrophysics (Nanjing University), Ministry of Education, Nanjing 210093, China}

\begin{abstract}
We investigate the footpoints of four erupted magnetic flux ropes (MFRs) that appear as sigmoidal hot channels prior to the eruptions in the Atmospheric Imaging Assembly high temperaure passbands. The simultaneous Helioseismic and Magnetic Imager observations disclose that one footpoint of the MFRs originates in the penumbra or penumbra edge with a stronger magnetic field, while the other in the moss region with a weaker magnetic field. The significant deviation of the axis of the MFRs from the main polarity inversion lines and associated filaments suggests that the MFRs have ascended to a high altitude, thus being distinguishable from the source sigmoidal ARs. The more interesting thing is that, with the eruption of the MFRs, the average inclination angle and direct current at the footpoints with stronger magnetic field tend to decrease, which is suggestive of a straightening and untwisting of the magnetic field in the MFR legs. Moreover, the associated flare ribbons also display an interesting evolution. They initially appear as sporadical brightenings at the two footpoints of and in the regions below the MFRs and then quickly extend to two slender sheared J-shaped ribbons with the two hooks corresponding to the two ends of the MFRs. Finally, the straight parts of the two ribbons separate from each other, evolving into two widened parallel ones. These features mostly conforms to and supports the recently proposed three-dimensional standard CME/flare model, i.e., the twisted MFR eruption stretches and leads to the reconnection of the overlying field that transits from a strong to weak shear with the increasing height.
\end{abstract}

Online-only material: animations, color figures

\section{Introduction}
Coronal mass ejections (CMEs) and flares are among the most spectacular explosive phenomena in our solar system. They are able to release various relativistic particles and eject a large quantity of plasma and magnetic flux with a velocity of hundreds of km s$^{-1}$, even up to 3000 km s$^{-1}$ \citep{yashiro04}, into the interplanetary space. When arriving at the Earth, these energetic particles and high-speed magnetized plasma may interact with the magnetosphere and ionosphere, and seriously affect the safety of human high-tech activities in the outer space such as disrupting communications, overloading power grids, presenting a hazard to astronauts, and so on \citep{webb94,liuying14}.

In the past decades, theoretical solar physicists proposed that a magnetic flux rope (MFR), a coherent magnetic structure with the field lines wrapping around the central axis more than once, is the fundamental structure in the CME/flare dynamical process \citep[e.g.,][]{shibata95,chenj96,titov99,chen11_review,vourlidas13}. \citet{forbes91} studied a cylindrical MFR and found that such helical structure can suddenly loss its equilibrium and rapidly erupt upward. \cite{kliem06} considered a freely expanded current torus. With a detailed analysis, they concluded that the MFR can experience a nonlinear expansion if the background magnetic field that constrains the MFR declines fast enough. Such an expansion instability is also named as torus instability. Following the work of \cite{kliem06}, \citet{olmedo10} further studied a more realistic situation, in which two footpoints of the MFR are line-tied to the photosphere, well resembling the pre-eruption state of the MFR. They pointed out that the threshold of torus instability not only depends on the decline of the background field but also on the geometrical circularity of the MFR: a ratio of the MFR arc length above the photosphere to its circumference. Besides torus instability, kink instability can also initiate an eruption. When the twist of the MFR exceeds a critical value, the axis of the MFR will rapidly deform, transforming part of the twist into writhe \citep{torok04,fan07}. Meanwhile, the height of the MFR quickly ascends, triggering the MFR eruption \citep[e.g.,][]{ji03,guo10_index,srivastava10,cheng14_kink,yanxl14}. 

In addition to ideal MHD instabilities, reconnection-associated mechanisms such as tether-cutting \citep{moore01} and breakout reconnection \citep{antiochos99,chen00,karpen12} are also proposed to interpret the initiation of the MFR. By contrast, these two models do not require that the MFR has to exist prior to the eruption. In the tether-cutting model, the erupted magnetic field initially is comprised of two groups of sheared arcades. The reconnection between the sheared arcades will form a helical magnetic field, i.e., the nascent MFR \citep[e.g.,][]{liuchang_13,chenhd14}. An upward Lorentz force is simultaneously generated to initiate the MFR. On the contrary, the breakout model consists of a quadrupole field configuration with a null point situated at between the central arcade and the overlying large-scale field. Once the reconnection near the null point commences, the overlying field will be opened, leading to the reduction of the downward tension and thus the rise of the central arcade \citep[e.g.,][]{zuccarello08,shenyuandeng12,lynch13}. 

No matter which model applies, the flare reconnection below the erupted configuration can be switched on, which rapidly produces the poloidal flux to accelerate the eruption of the MFR. At the same time, the fast reconnection generates energetic electrons, which stream down along newly reconnected field lines and heat the chromosphere, mapping two elongated ribbon-like structures and casuing the flare emission in various electromagnetic passbands \citep{priest02}. Different from those ideal MHD mechanisms that suppose a pre-existing MFR, the tether-cutting reconnection forms the nascent MFR in the slow rise phase, which further grows up in the acceleration phase, while the breakout reconnection does not build up the MFR until in the acceleration phase. 

Owing to the important role of the MFR in the CME/flare dynamics, the solar community has been searching for the observational evidence of the MFR. In solar active regions (ARs), forward or reversed sigmoidal emission pattern in extreme ultraviolet (EUV) and soft X-ray (SXR) passbands is often seen as a progenitor of the CME \citep[e.g.,][]{rust96,canfield99,sterling00,liuc07,mckenzie08,liur10,savcheva12a}, thus being considered as evidence of the MFR preexisting in the corona. On the other hand, filaments correspond to a collection of relatively dense and cool plasma suspended at dips of a helical structure; thus, filaments also serve as an indicator of the MFR \citep{mackay10}. If there are no filament materials, the helical MFR structure probably appears as dark filament channels along the polarity inversion line (PIL) as been visible in the EUV passbands \citep{vanballegooijen98,chenpf14}. When rotating to the solar limb, the cross section of filament channels will manifest as dark cavities. The frequently observed spinning motions \citep{low95_apj,gibson04,wangym10,lixing12}, bright ring \citep{dove11}, and ``lagomorphic" structure of linear polarization \citep{bak-steslicka13} in the cavities, are even thought to be direct evidence of the existence of helical magnetic field. Moreover, the descending motion of filament materials along a helical trajectory is also argued to be strong evidence of a helical structure \citep[e.g.,][]{liting13,cheng14_tracking,yang14,zhangjun15}. The intrinsic relationship between filaments and MFRs has further been supported by various case studies, in which the dips of the twisted field lines reproduced by the non-linear force free field (NLFFF) extrapolation basically conform with the filament locations \citep[e.g.,][]{guo10_filament,savcheva12a,suyingna12,cheng13_double,inoue13,jiang14_nlfff,jiang14_filament,jiang16a,yanxl15}.  
 
Recently, \citet{cheng11_fluxrope} presented unambiguous evidence of an MFR being formed in the lower corona \citep[also see;][]{song14_formaion,song14}. It manifests as an EUV blob in the high temperature passbands of the Atmospheric Imaging Assembly \citep[AIA;][]{lemen12}. In the other low temperature passbands, however, it appears as a dark cavity. Subsequently, \citet{zhang12} and \citet{cheng13_driver} unveiled that, if seen off the solar limb, the MFR will take on a coherent channel-like hot structure that exists prior to the eruption \citep[also see;][]{patsourakos13,lileping13,liting13_homologous,tripathi13,chintzoglou15,joshi15,cheng15_iris}. Once the hot channel takes off, it gradually evolves to a semicircular loop-like structure in the early rise phase. Afterwards, the loop-like structure quickly drives the formation of the CME and gives rise to the flare emission simultaneously. Furthermore, through a detailed case study, \citet{cheng14_tracking} identified that the hot channel can evolve smoothly from the inner into the outer corona by retaining its coherence, whose morphology coincides with the CME cavity in the white-light images. Moreover, \citet{cheng14_kink} and \citet{chenbin14} documented that the hot channel is initially cospatial with but later on separates from the associated prominence. At the later time, the prominence only occupies the lower part of the hot channel spatially. \citet{nindos15} further made a statistical study on the appearance frequency of the hot structures and found that almost half of major eruptive flares contain a hot blob or channel-like configuration. All of these studies support the arguement that the EUV hot blobs and channels are actually the helical MFRs that exist prior to the eruption.

Therefore, it can be said that the EUV hot channels can now be regarded as the most promising evidence of the MFRs. However, previous studies on the MFRs are mostly restricted to their morphology or kinematics, while no attention is paid to the properties of their footpoints. This is mainly due to that it is extremely difficult to find out appropriate hot channels events, in which the footpoints can be identified unambiguously. In the current study, by inspecting tens of solar eruption events, we find out four hot channels events which are well observed and have the discernable footpoints. Furthermore, we detailedly investigate various characteristics at the footpoints of the MFRs, as well as their relation to the flare ribbons. On the other hand, \cite{aulanier12} and \cite{janvier13} recently extended the two-dimensional (2D) standard CME/flare model \citep[CSHKP;][]{carmichael64,sturrock66,hirayama74,kopp76} to 3D, which predicts the vertical straightening of the inner legs of the CME and the transfer of the magnetic shear from strong to weak after the eruption. Thus, the revealed observational characteristics can also be used to compare and validate the 3D CME/flare model. In Section 2, we introduce the observing instruments, followed by the event selection in Section 3. In Section 4, we display the results. Finally, we give a summary and discussions in Section 5.

\section{Instruments}
The data sets used in this study are primarily from \textit{Solar Dynamics Observatory} \citep[\textit{SDO};][]{pesnell12}. The AIA on board \textit{SDO} provides images of the solar corona that are acquired almost simultaneously at temperatures ranging from 0.06 MK to 20 MK through 9 passbands (7 EUV and 2 UV passbands) with a temporal cadence of 12 s or 24 s and spatial resolution of 1.2\arcsec. The Helioseismic and Magnetic Imager \citep[HMI;][]{schou12} also on board \textit{SDO} observes the photosphere at 6173 {\AA} and provide light-of-sight magnetograms of the full disk with a temporal cadence of 45 s and spatial resolution of 1.0\arcsec. The vector magnetograms are provided by HMI Active Region Patches \citep[HARPs;][]{bobra14} product, in which the 180$^{\circ}$ ambiguity in the transverse component has been removed. Moreover, we take advantage of the Large Angle and Spectrometric Coronagraph \citep[LASCO;][]{brueckner95} on board \textit{Solar and Heliospheric Observatory} (\textit{SOHO}) and \textit{GOES} to inspect CMEs and monitor the soft X-ray (SXR) 1--8 {\AA} flux of the associated flares, respectively.

\section{Event Selection}
In order to investigate the footpoints of MFRs, we search for appropriate events from the ISEST Event List\footnote{http://solar.gmu.edu/heliophysics/index.php/The\_ISEST\_Event\_List}. The selection criteria are: (1) A hot channel-like structure, which is believed to be the proxy of the MFR \citep{zhang12,cheng13_driver}, can be clearly seen prior to the eruption in the AIA 131 and 94 {\AA} passbands; (2) The source regions of hot channels are near the disk center so as to avoid the projection effect of magnetic field measurement; (3) The corresponding interplanetary CMEs have clear magnetic cloud structures when arriving at the Earth and produce geomagnetic storms; and (4) The magnitude of the associated flares is $\ge$M1.0, which ensures that the footpoints of the MFRs have enough intensity so that their locations can be tracked precisely. 

\begin{figure}
\center {\includegraphics[width=12cm]{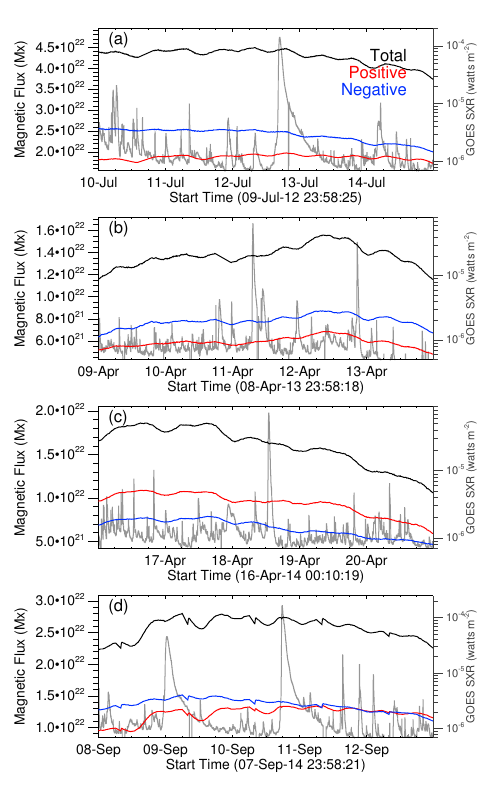}}
\caption{Temporal evolution of the positive flux (blue), negative flux (red), and total unsigned flux (black) in the four MFR-hosting ARs. The gray curves show the \textit{GOES} SXR 1--8 {\AA} fluxes, in which the maximum peak in each panel corresponds to the flare caused by the erupted hot channels.}
\label{goes}
\end{figure}

Four events are found to fit the above criteria as shown in Table \ref{tb1}. One can see that the four flares have a long rise time ($>$20 minutes) and a large magnitude ($>$M6.0), and that the four CMEs are full halo ones with velocities of $>$800 km s$^{-1}$. Figure \ref{goes} shows the long-term evolution of the positive and negative magnetic fluxes in their source ARs. We find that all eruptions appear in the decay phase of the ARs but at their early stage. Note that, for the SOL2013-04-11T event, the increase of the positive and negative fluxes is due to the emergence of a second bipolar that appears in the northwest of the main bipolar. Moreover, we can see that in a long period, at least one day, before and after the eruption, there is no other big event taking place, which shows that the magnetic environment of the four eruptions is not influenced by other major eruptions.

\section{Results}
\subsection{Morphology of the MFRs}
\begin{figure}
\center {\includegraphics[width=11.9cm]{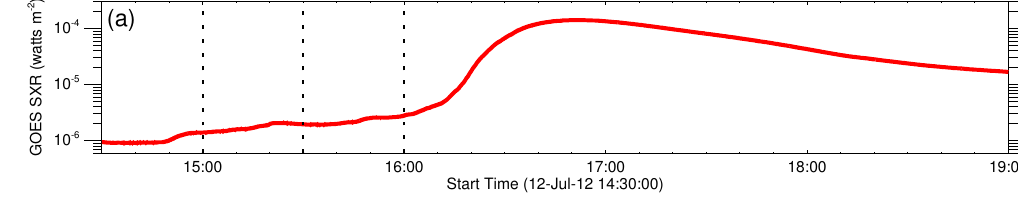}\vspace{-0.03\textwidth}}
\center {\includegraphics[width=12cm]{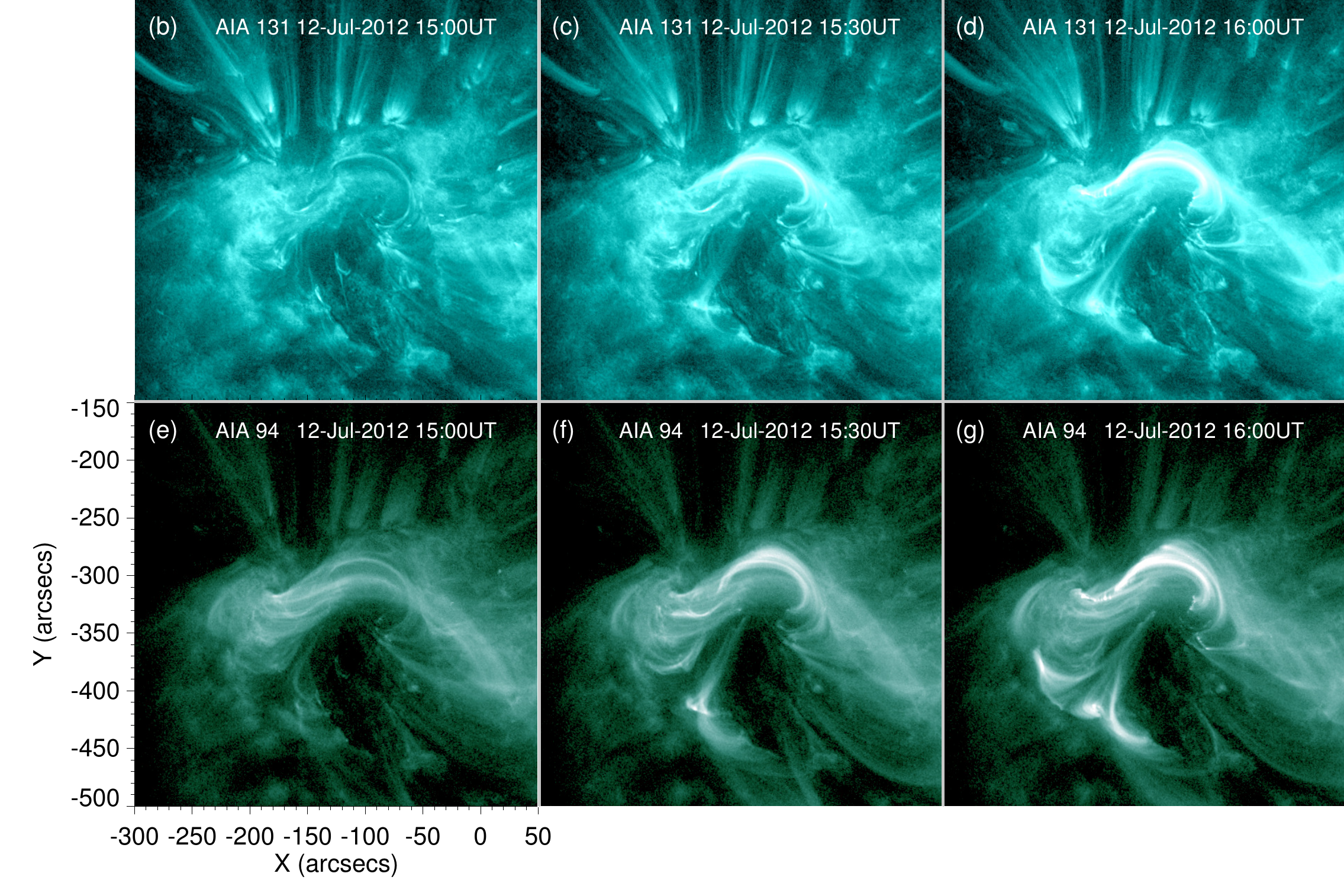}}
\caption{(a) \textit{GOES} SXR 1--8 {\AA} flux of the SOL2012-07-12T flare. (b)--(g) \textit{SDO}/AIA 131 {\AA} ($\sim$0.4, 11, 16 MK) and 94 {\AA} ($\sim$1.1 and 6.3 MK) images showing the early evolution of the SOL2012-07-12T hot channel at three times. The corresponding times are indicated by three vertical dashed lines in panel a.}
\label{0712_erupt}
\end{figure}

\begin{figure}
\center {\includegraphics[width=11.9cm]{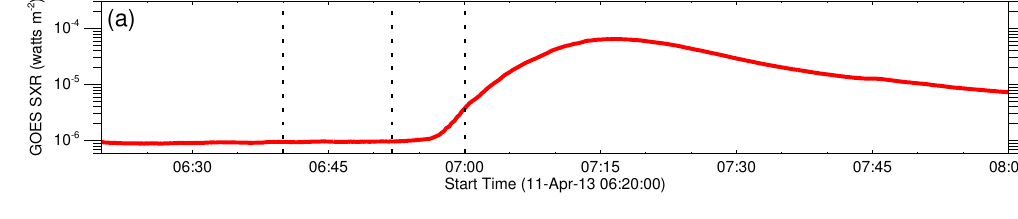}\vspace{-0.03\textwidth}}
\center {\includegraphics[width=12cm]{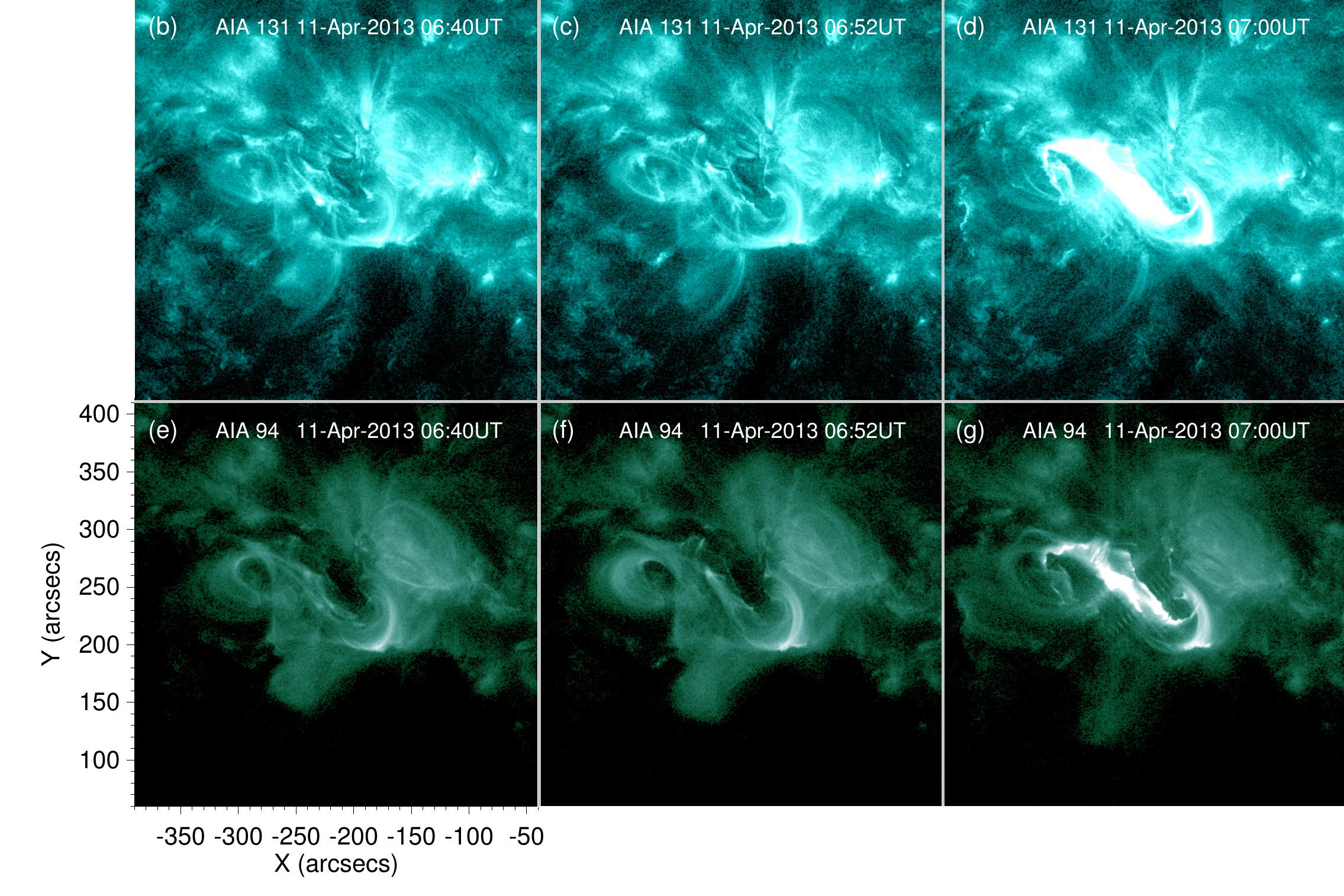}}
\caption{Same as Figure \ref{0712_erupt} but for the SOL2013-04-11T hot channel.}
\label{0411_erupt}
\end{figure}

\begin{figure}
\center {\includegraphics[width=11.9cm]{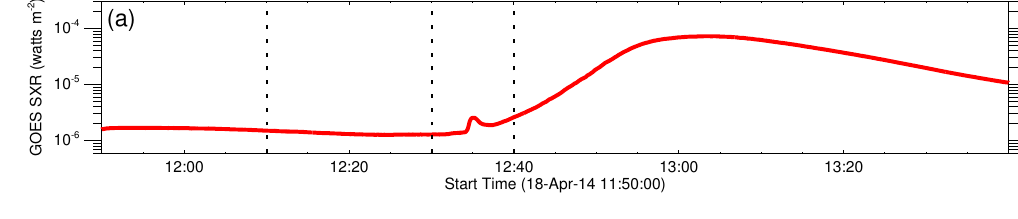}\vspace{-0.03\textwidth}}
\center {\includegraphics[width=12cm]{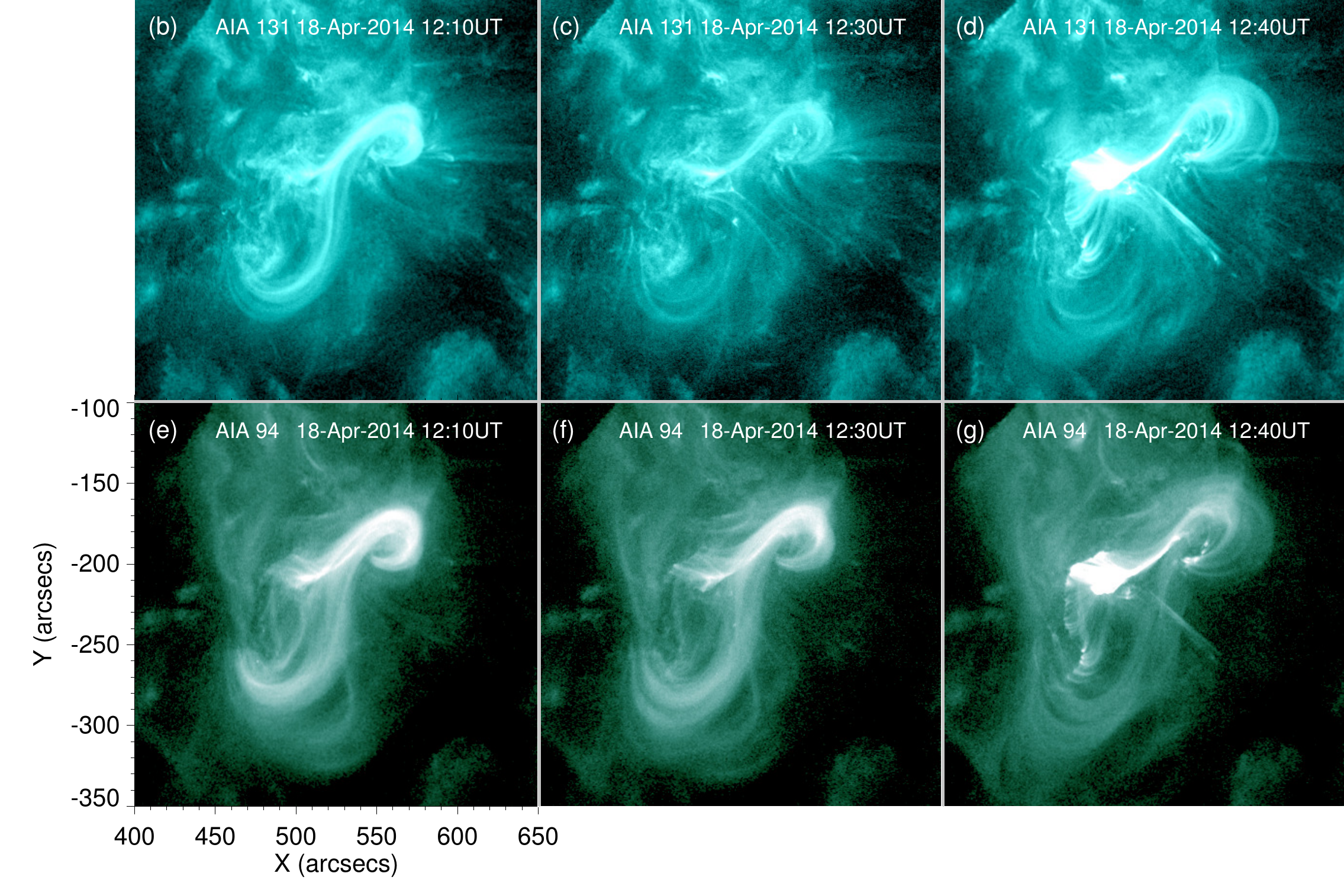}}
\caption{Same as Figure \ref{0712_erupt} but for the SOL2014-04-18T hot channel.}
\label{0418_erupt}
\end{figure}

\begin{figure}
\center {\includegraphics[width=11.9cm]{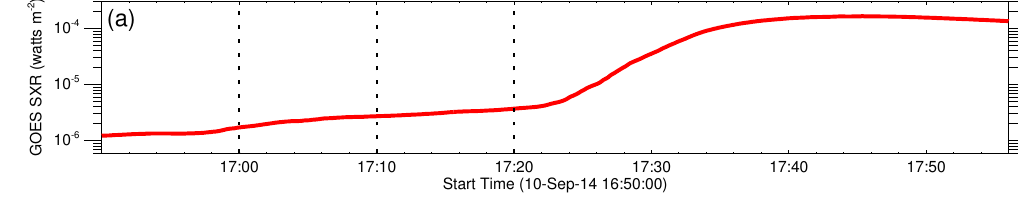}\vspace{-0.03\textwidth}}
\center {\includegraphics[width=12cm]{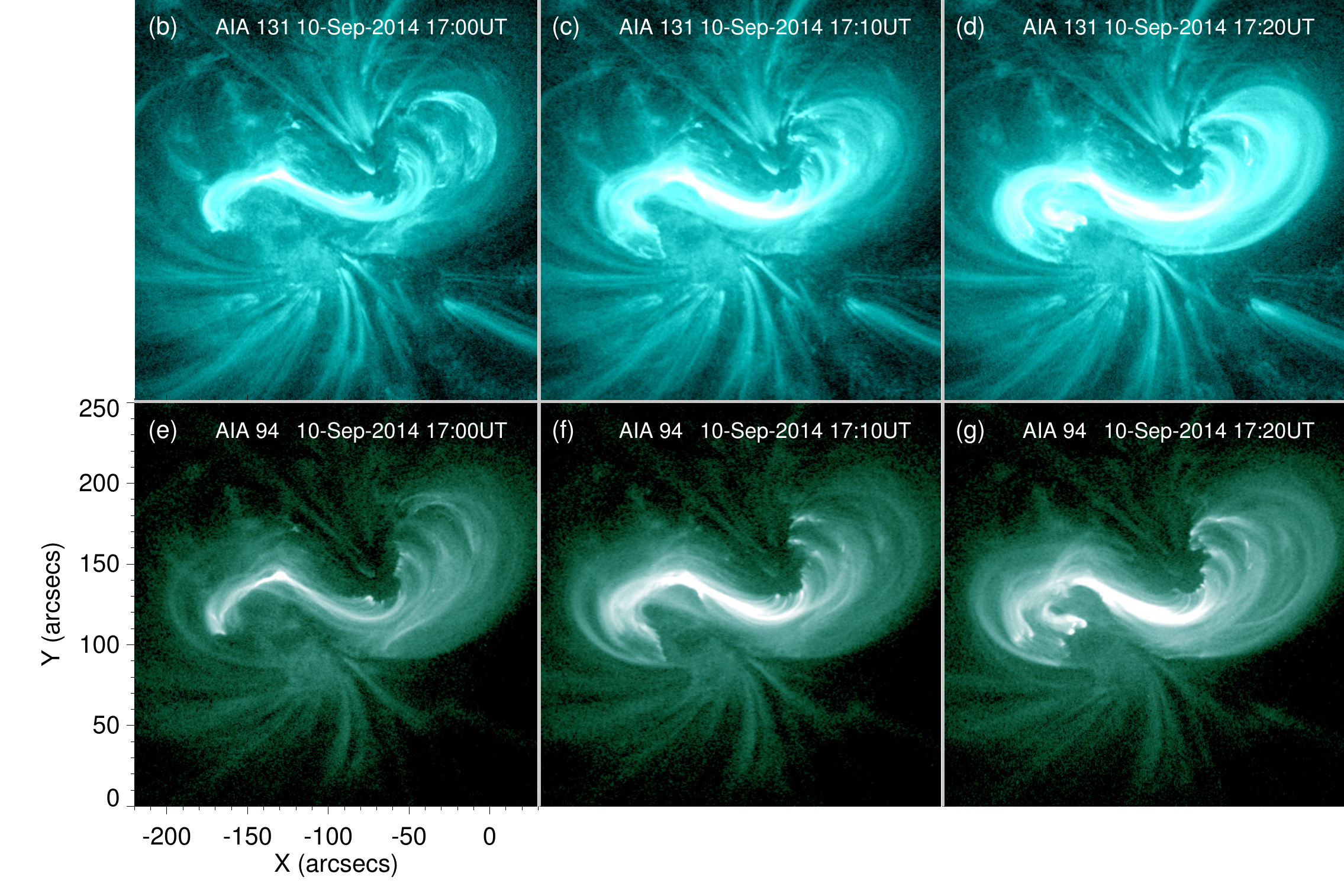}}
\caption{Same as Figure \ref{0712_erupt} but for the SOL2014-09-10T hot channel.}
\label{0910_erupt}
\end{figure}

\begin{figure}
\center {\includegraphics[width=15cm]{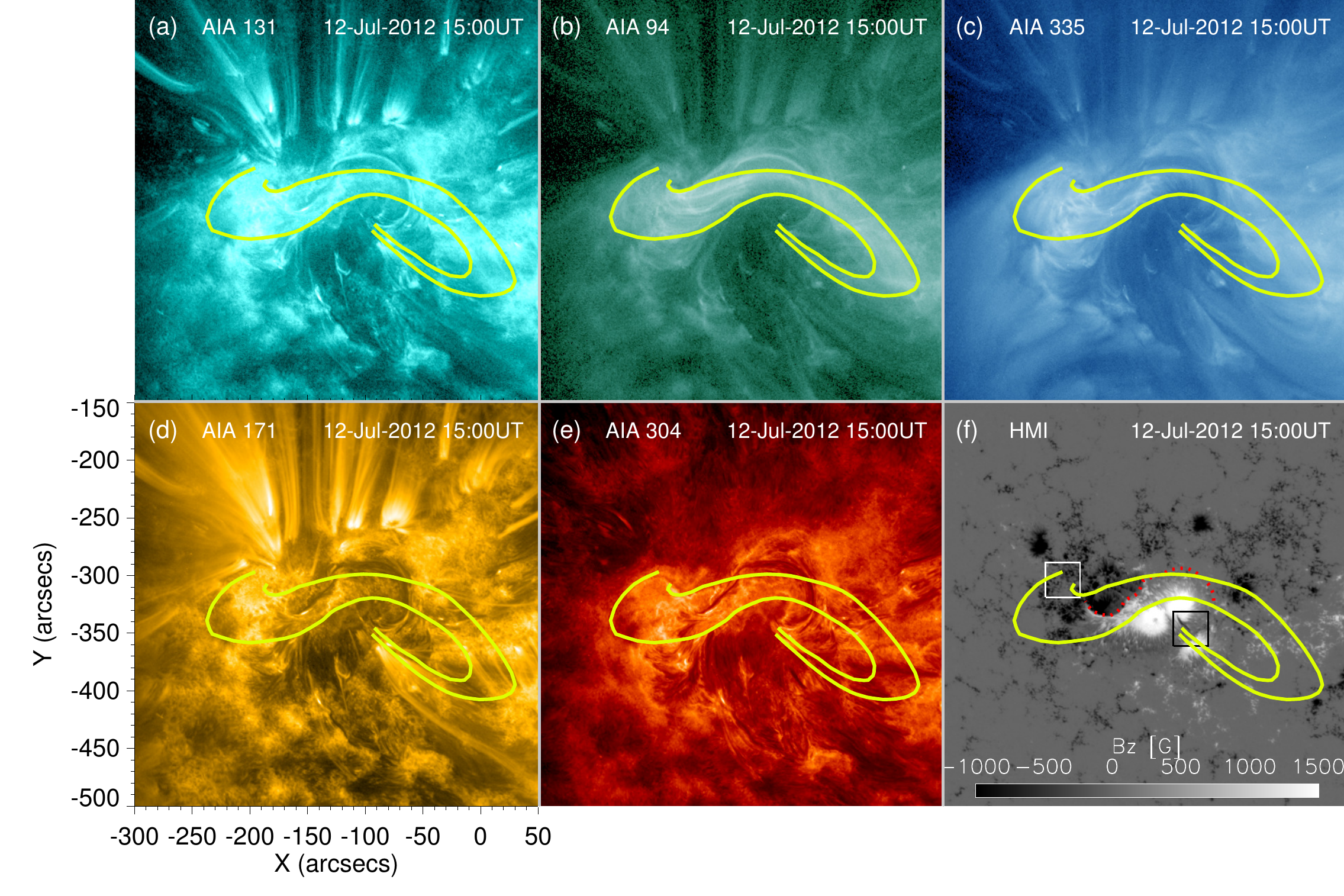}}
\caption{(a)--(e) \textit{SDO}/AIA 131 {\AA} ($\sim$0.4, 11, 16 MK), 94 {\AA} ($\sim$1.1 and 6.3 MK), 335 {\AA} ($\sim$2.5 MK), 171 {\AA} ($\sim$0.6 MK), and 304 {\AA} ($\sim$0.05 MK) images of the AR producing the SOL2012-07-12T event. Two yellow curves roughly depict the boundaries of the SOL2012-07-12T hot channel. (f) \textit{SDO}/HMI light-of-sight magnetogram overlaid with the hot channel (yellow) and the main PIL (red). The white (black) box with the FOV of 30\arcsec$\times$30\arcsec denotes the region of the negative (positive) footpoints of the hot channel.}
(Animation of this figure is available in the online journal.)

\label{0712}
\end{figure}

\begin{figure}
\center {\includegraphics[width=15cm]{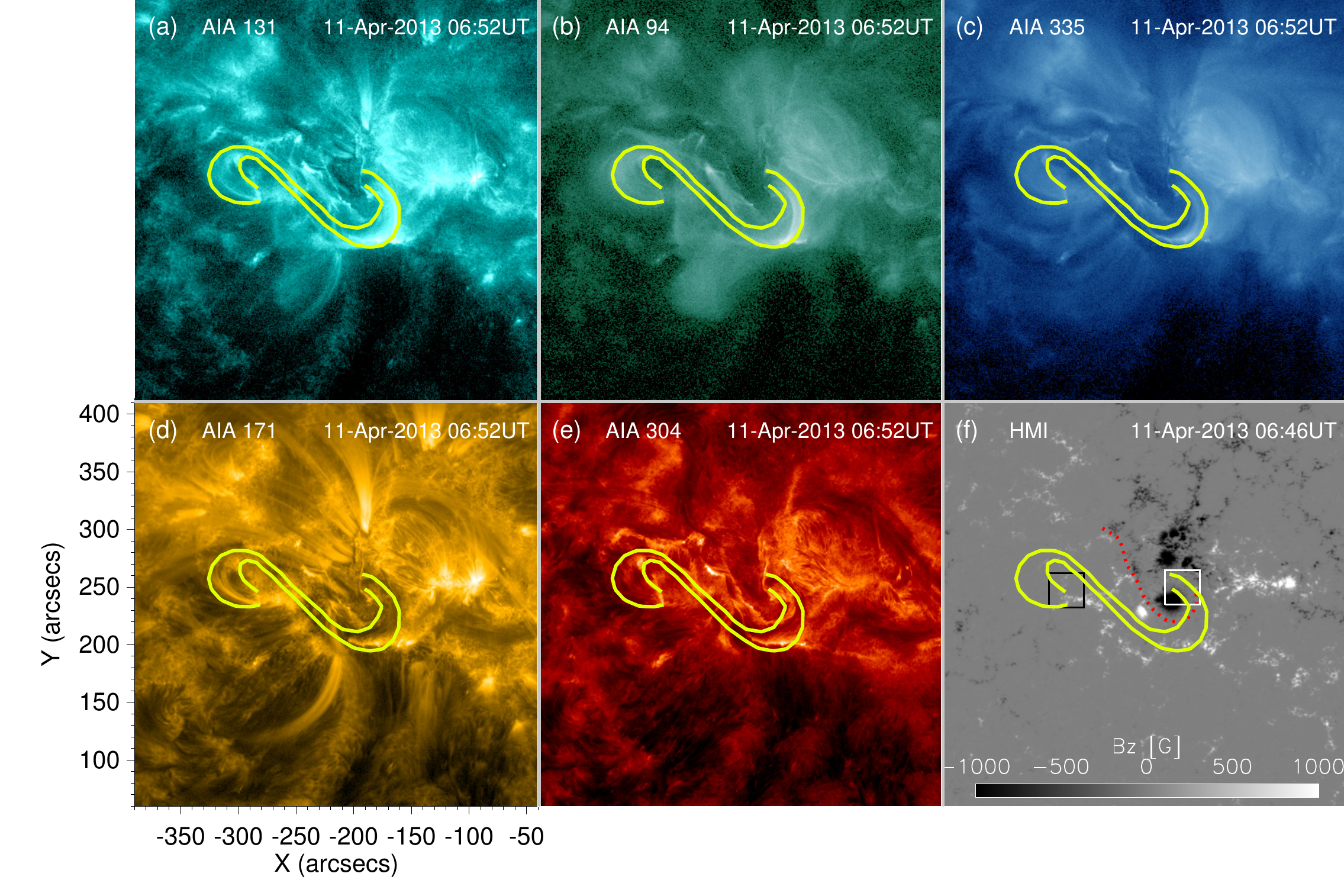}}
\caption{Same as Figure \ref{0712} but for the SOL2013-04-11T event.}
(Animation of this figure is available in the online journal.)

\label{0411}
\end{figure}

\begin{figure}
\center {\includegraphics[width=15cm]{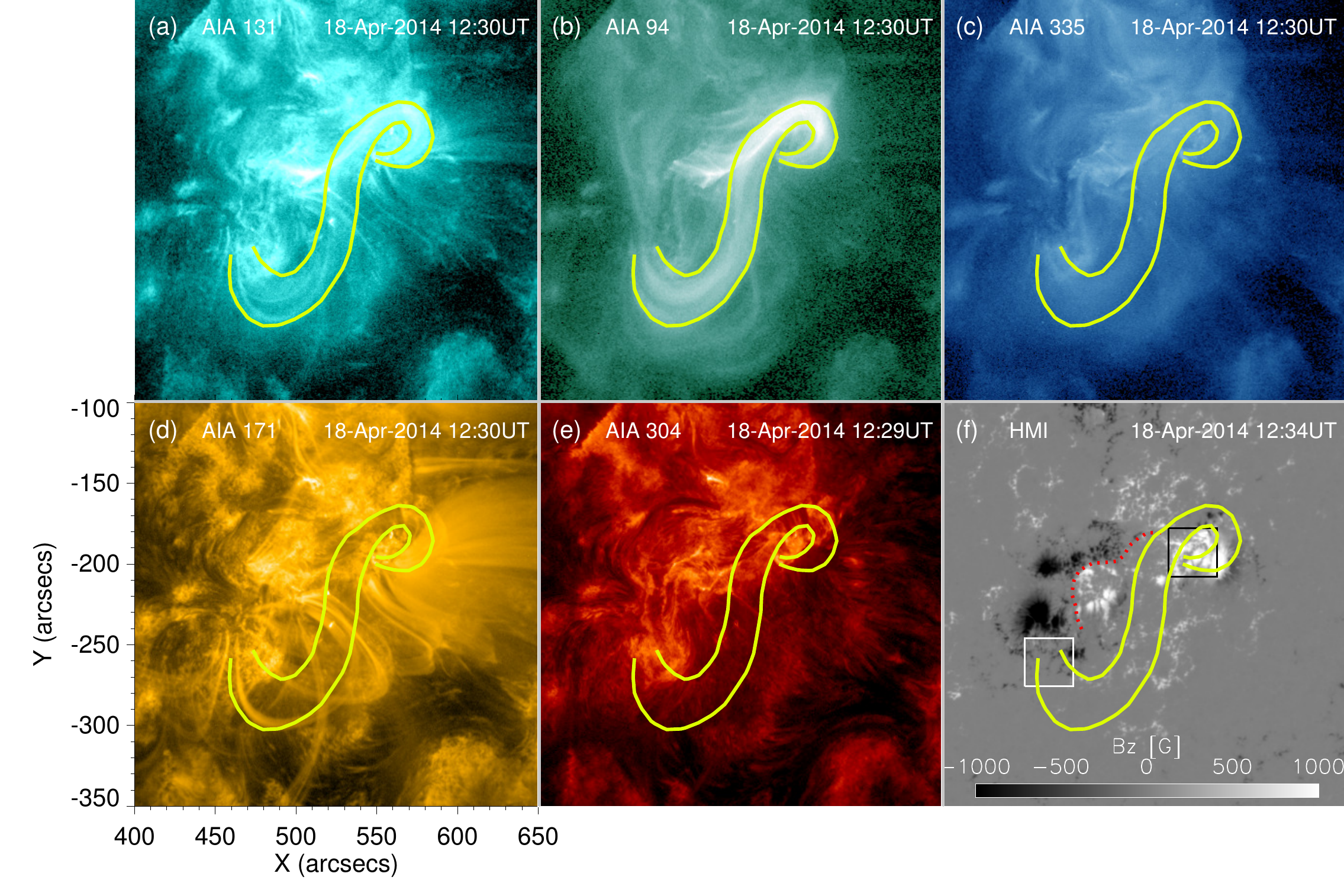}}
\caption{Same as Figure \ref{0712} but for the SOL2014-04-18T event.}
(Animation of this figure is available in the online journal.)

\label{0418}
\end{figure}

\begin{figure}
\center {\includegraphics[width=15cm]{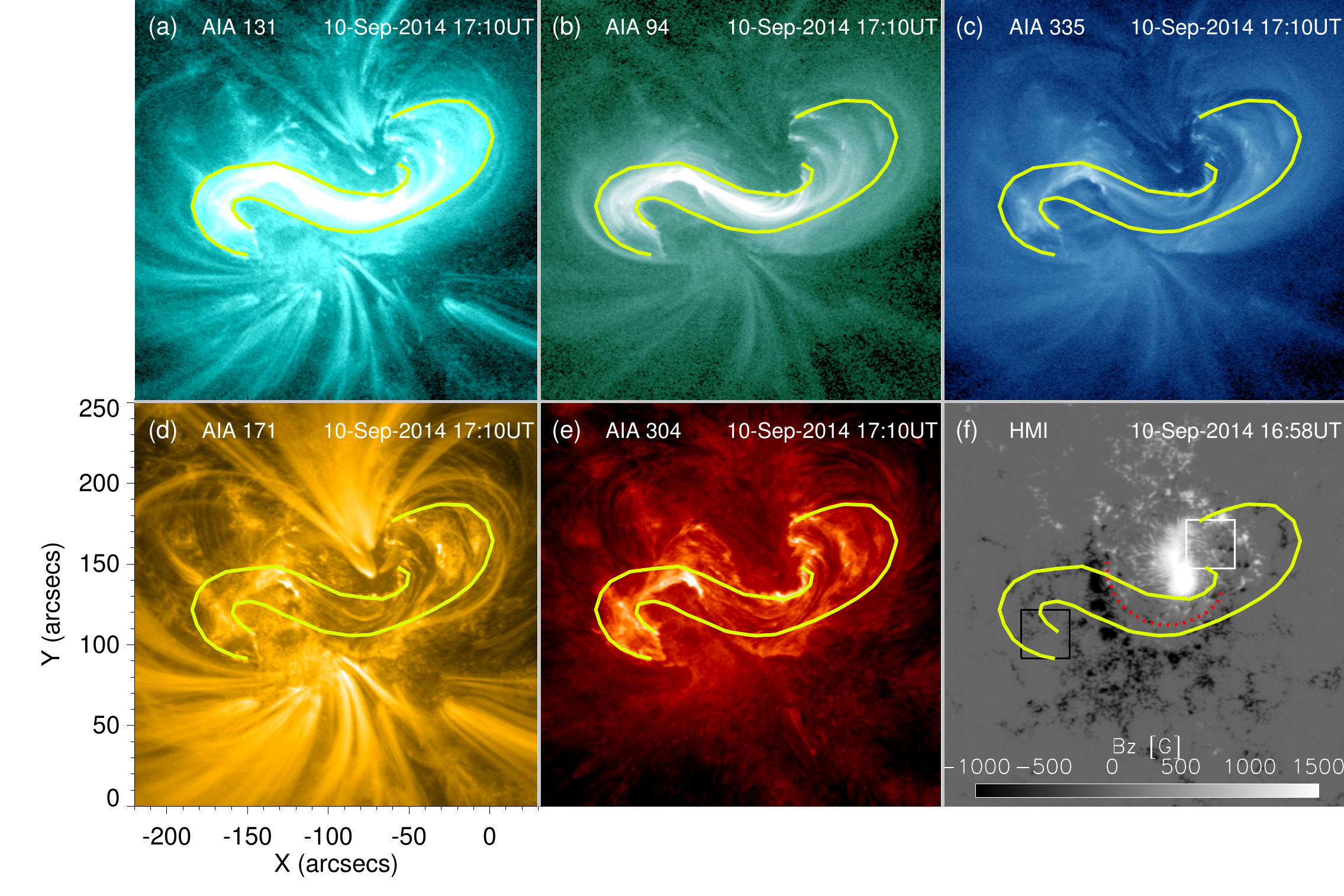}}
\caption{Same as Figure \ref{0712} but for the SOL2014-09-10T event. }
(Animation of this figure is available in the online journal.)
\label{0910}
\end{figure}

We find that prior to the eruption, the most apparent feature appearing in the four ARs is a curved cylinder or writhed channel-like structure as seen in the AIA 131 {\AA} and 94 {\AA} images. From Figure \ref{0712_erupt}--\ref{0910}, one can see that all of the channels take on a forward (SOL2012-07-12T and SOL2014-04-18T events) or reversed (SOL2013-04-11T and SOL2014-09-10T events) sigmoidal shape when approaching the eruption, at least in its early phase. Whereas in the AIA 171 {\AA} passband, the ARs mainly display the overlying loops that straddle over and are almost perpendicular to the channels. The channels most likely contain plasmas of high temperatures since they only appear in the AIA high temperature passbands and are almost invisible in other low temperature passbands. Actually, the detailed differential emission measure analyses, which have been applied to the SOL2012-07-12T and SOL2013-04-11T events by \cite{cheng14_formation} and \cite{vemareddy14}, respectively, proved that the channels do have a high temperature of $\ge$8 MK. Because of the high temperature property, \cite{zhang12} and \cite{cheng13_driver} named the curved cylinder or writhed channel structures as EUV hot channels.

It is worth noting that, although the ARs often take on a sigmoidal shape in the EUV and/or soft X-ray images, it does not mean that the magnetic field lines there must be continuously sigmoidal threads \citep{titov99,kliem04,valori10,schmieder15,jiang16}. Alternatively, there could be two groups of sheared arcades, making up a sigmoidal shape as a whole \citep[e.g.,][]{rust96,canfield99,mckenzie08,liur10}. However, the latter explanation is not appropriate for the four hot channels studied here, because they are mostly comprised of continuously sigmoidal threads for the reasons stated in detail below.

As for the SOL2012-07-12T case, \cite{cheng14_formation} studied the formation of the hot channel-like MFR in detail and found that the twisted field lines have been well built up a half day before the eruption. Furthermore, they found that, two hours before the eruption, the whole sigmoid actually is a double-decker MFR structure; the low-lying one is manifested by the filament and the high-lying one is evidenced by the strongly writhed channel-like structure (Figure \ref{0712_erupt}). With the flare beginning, it is clearly seen that the high-lying channel detach from the sigmoidal AR with the middle part moving toward the south. The high-lying hot channel being an MFR is also proved by \cite{dudik14} with the MHD simulation, in which they showed that the high-lying sigmoidal threads and their evolution are well reproduced by an unstable MFR and its eruption process.

Figure \ref{0411_erupt} presents the early evolution of the SOL2013-04-11T sigmoid (also see the attached online movie of Figure \ref{0411}). One can see that a reversed sigmoid first appears in the core field of the source AR. From $\sim$06:45 UT, part of the S-shaped field lines start to detach from the sigmoid and then slowly rise up with the middle part having the fastest speed. After ten minutes, the S-shaped field lines show an M-shape, which quickly evolves into a semicircular shape. This similar evolution behavior to the SOL2011-03-08T event analyzed by \cite{zhang12} and \cite{cheng13_driver} strongly suggests that the erupted S-shaped threads are actually the MFR \citep[also see,][]{vemareddy14}.

For the SOL2014-04-18T event, the forward sigmoidal channel appears as early as one hour before the eruption. Thanks to the high resolution of the AIA, we can see that the channel structure mainly consists of a group of coherently sigmoidal threads (Figure \ref{0418_erupt}). Through analyzing the IRIS spectral lines formed in the transition region and the chromosphere and the HMI magnetic field in the photosphere, \citet{cheng15_iris} concluded that the formation of the continuously sigmoidal threads in the hot channel is mainly via tether-cutting reconnection in the chromosphere, which gradually transforms the sheared arcades into twisted field lines prior to the eruption. \cite{joshi15} also studied this event but mostly focused on the flare. They found that the beginning of the flare is triggered by the eruption of the pre-existing sigmoidal hot channel, which is also identified as the MFR (see their Figure 3).
 
The SOL2014-09-10T hot channel is slightly different from the above three ones. The sigmoidal emission pattern of the source AR is mostly a result of two groups of sheared arcades, which manifest a sigmoid as a whole (Figure \ref{0910_erupt}). However, through further checking the property of the magnetic field at the sigmoid center in the period of five hours before the eruption, \citet{cheng15_iris} found that magnetic reconnection has been playing a role in transforming part of sheared arcades into longer threads. Recently, using Grad-Rubin method, \citet{zhaojie16} extrapolated the 3D NLFFF field of the AR two hours before the eruption and also found that some twisted field lines are comparable with the sigmoidal loops with, though, a small twist number ($\sim$1.5). Actually, in the early phase of the eruption, we do see a set of continuously sigmoidal threads that detach from the source sigmoidal AR and then erupt outward.

By carefully inspecting the orientation of the four hot channels, we also find that it has some deviation from that of the main PILs of the ARs (red dashed lines in Figure \ref{0712}f--\ref{0910}f). It indicates that the hot channels are most likely located in the corona with a large separation in altitude from the PILs of the source ARs in the photosphere. Owing to the projection effect, the SOL2013-04-11T (SOL2014-04-18T) hot channel in the ARs located at N07E13 (S20W34) deviates from the PIL toward the east (west). Moreover, in the events except the SOL2014-04-18T one, the filaments appear to lie nicely along the main PILs as seen in the AIA 304 {\AA} images. Interestingly, during the CME eruptions, the filaments do not erupt. It shows again that the erupted hot channels are magnetic structures different from the source sigmoids and filaments.

To sum up, we can say that, although the four sigmoids seen in the EUV images may be comprised of two groups of sheared arcades, the hot channels should be the MFRs since they have ascended to a certain height in the corona and thus separated from the sigmoids, at least in the slow rise phase. Taking advantage of the AIA 131 {\AA} and 94 {\AA} images, we identify and depict the boundaries of the four hot channel-like MFRs, as shown (yellow lines) in Figures \ref{0712}--\ref{0910}.

From Figure \ref{0712}b, \ref{0411}b, \ref{0418}b, \ref{0910}b, and Table \ref{tb2}, we find that all of hot channels have a similar footpoint separation distance. We introduce a parameter $R$, the ratio of the projected length of the hot channels to the footpoint separation, to roughly estimate the writhe of the axis of the hot channels. The projected lengths of the four channels, which actually are the lower limit of the real lengths, range from 183 to 312 Mm. They are much larger than the separation distances (67--80 Mm) of the two footpoints. It indicates that the hot channels not only detach from the PILs but also have a significant writhe. One can see that the SOL2012-07-12T and SOL2013-04-11T hot channels have a very large $R$, indicating a strong writhe, in addition to the twist in the threads of the hot channels (Figures \ref{0712} and \ref{0411}). By contrast, the SOL2014-04-18T and SOL2014-09-10T hot channels have a relatively small $R$, which is consistent with a weak writhe in the corresponding threads (Figures \ref{0418} and \ref{0910}).
   
\subsection{Overall Properties of Source Regions of the MFRs}
We first study the magnetic properties of the ARs that host the hot channels. Here we use the HARP cylindrical equal-area vector field data (Figures \ref{0712_vector}a, \ref{0411_vector}a, \ref{0418_vector}a, and \ref{0910_vector}a), in which the magnetic field, $\bf B$, has been remapped to a lambert cylindrical equal-area projection, i.e., the Heliocentric Earth Ecliptic Cartesian coordinate system, and includes projection correction and geometric modification \citep{sunxudong13}. Based on the Amp$\grave{e}$re's law, the vertical current density can be calculated through $j_{z}=(\nabla \times\mathbf{B})_z$/$\mu_{0}$, where $\mu_{0}$=4$\pi \times$10$^{-3}$ G m A$^{-1}$. Note that, in order to avoid the effect of noise in the weak fields, the calculation is only done for $B_z>$50 G. Moreover, assuming that the uncertainty in the transverse field $B_{t}$ is $\sim$100 G, following the formula $\delta j_{z}=\delta B_{t}/(\mu_{0}\delta x)$ \citep{gary95}, the error of $j_{z}$ is estimated to be of the order of 0.02 A m$^{-2}$. We also compare the values of $j_{z}$ calculated from the HARP data with that calculated from the vector data that are reverted with the UNNOFIT code \citep{janvier14_current}, which are shown to be almost the same.

Figures \ref{0712_vector}b, \ref{0411_vector}b, \ref{0418_vector}b, and \ref{0910_vector}b show the distribution of the vertical current density in the ARs. We find that the strongest current density always appears near the PILs, in particular for the SOL2012-07-12T and SOL2014-09-10T events. This is consistent with previous studies, which stated that magnetic free energy is mainly stored there \citep[e.g.,][]{sun12a}. Moreover, the strong current density at the PILs also indicates that magnetic reconnection probably occurs there, which converts the sheared arcades into the twist field so as to form the MFRs \citep[e.g.,][]{cheng14_formation,vemareddy14}. However, it should be noticed that those currents seem not to be related to the erupted hot channels. For example, in the SOL2012-07-12T, SOL2013-04-11T, and SOL2014-04-18T events, the filaments, which usually denote the locations with strong currents along the PILs, do not erupt following the eruption of the hot channels.

Figures \ref{0712_vector}c, \ref{0411_vector}c, \ref{0418_vector}c, and \ref{0910_vector}c show the distribution of the inclination angle in the ARs. One can see that the inclination angle increases from the sunspot center to outward and reaches its maximum in the penumbra. This conforms with the canopy-type configuration of ARs. Moreover, along the PILs, the inclination angle is also large, which is, however, due to the shear of the magnetic field, as shown in Figures \ref{0712_vector}a, \ref{0411_vector}a, \ref{0418_vector}a, and \ref{0910_vector}a. Due to the oppositely directed movements of magnetic footpoints (see attached online movies), the original potential field become more and more sheared, thus causing the large inclination angle.

Previous studies found that, in well-isolated ARs, the net current $I_n$ (the sum of direct current $I_d$ and return current $I_r$) over the positive polarity is qual to that over the negative one, i.e., the total current over the whole ARs is close to zero \citep{wheatland00,georgoulis12}. However, whether the net current over one polarity is also close to zero is debated. Here, we also tentatively calculate the net current over one polarity in the four ARs. Note that, for the forward sigmoidal ARs, which usually have a right-handed twist, the direct (return) current is parallel (antiparallel) with the direction of the magnetic field; while for the reversed sigmoidal ARs that usually have a left-handed twist, the direct (return) current is antiparallel (parallel) with the direction of the magnetic field \citep{rust96,green07,torok14}. The results are shown in Table \ref{tb2}, from which one can see that for the SOL2012-07-12T and SOL2014-09-10T cases, the net current significantly deviates from zero. While for the SOL2013-04-11T and SOL2014-04-18T cases, the net current almost vanishes.
  
\subsection{Magnetic Properties of Footpoints of the MFRs}
We consider the two ends of the hot channels as the footpoints of the MFRs (From Figures \ref{0712}f, \ref{0411}f, \ref{0418}f, and \ref{0910}f). We find that the two footpoints do not overlap the two ends of the main PILs. The fact is that, for all of hot channels, one footpoint comes from the moss region near the following sunspot, while the other one stems from the penumbra (SOL2014-04-18T) or penumbra edge (SOL2012-07-12T, SOL2013-04-11T, and SOL2014-09-10T) of the preceding sunspot. In the moss region, the magnetic field is weaker and the inclination angle is smaller; while in the penumbra and penumbra edge, the magnetic field is stronger and the inclination angle is larger.

It is noticed that the strong currents appear not only near the PILs, but also at the footpoints of the hot channels with the relatively strong magnetic field (Figures \ref{0712_vector}b, \ref{0411_vector}b, \ref{0418_vector}b, and \ref{0910_vector}b), the latter of which may be related to the currents of the MFRs. For the SOL2012-07-12T, SOL2014-04-18T, and SOL2014-09-10T (SOL2013-04-11T) events, at the positive (negative) footpoints, there always appears a dominated direct current. Moreover, we find that the distribution of the direct current takes on a curled shape on a large scale, which is surrounded by some sporadic return current, in particular for the SOL2012-07-12T (Figure \ref{0712_vector}b and \ref{0712_vector}h), SOL2013-04-11T (Figure \ref{0411_vector}b and \ref{0411_vector}h), and SOL2014-09-10T events (Figure \ref{0910_vector}b and \ref{0910_vector}h) \citep[also see,][]{janvier14_current}. 

\begin{figure}
\center {\includegraphics[width=14.9cm]{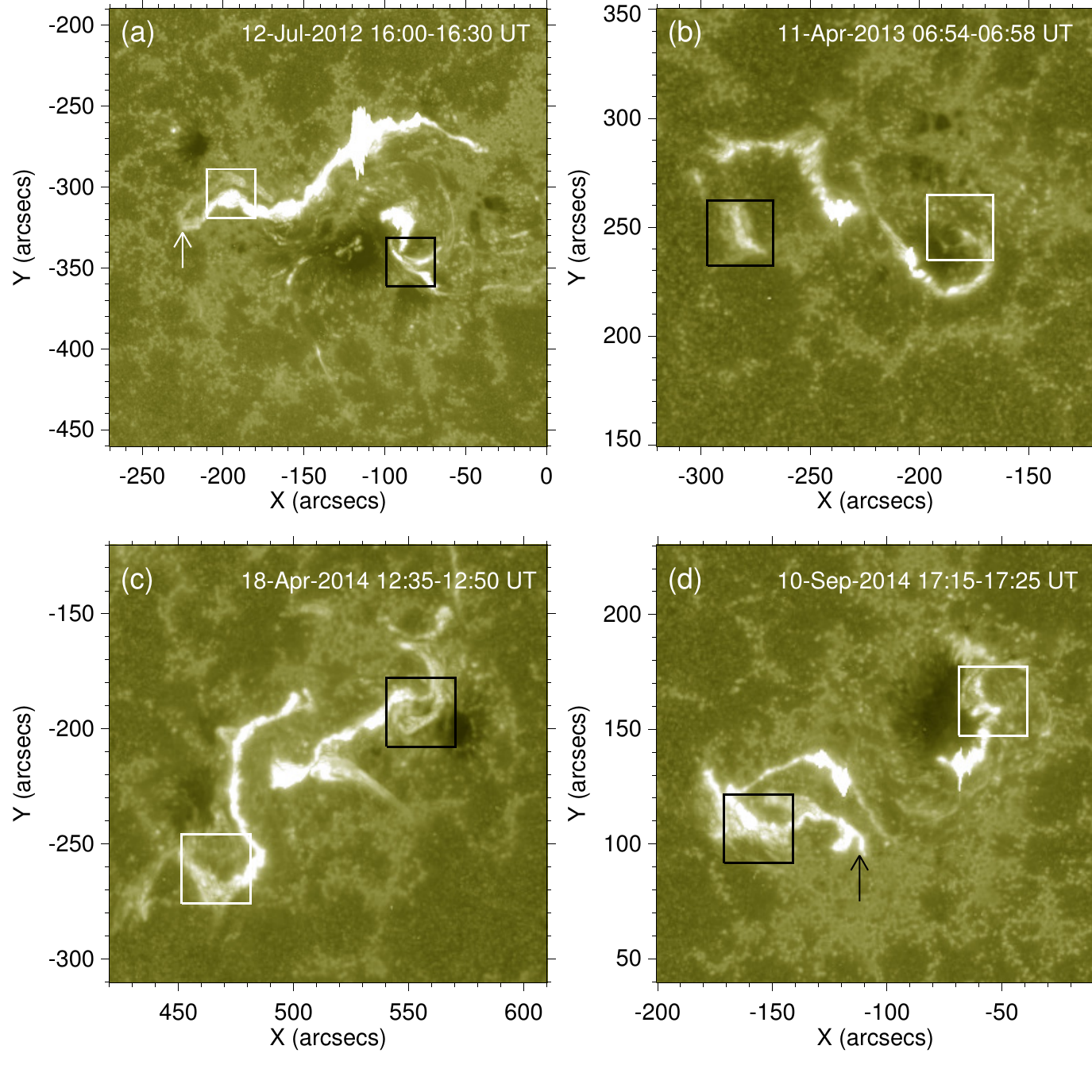}}
\caption{Synthesized flare ribbons that are derived through summing up the SDO/AIA 1600 {\AA} images in the time range as shown in the top of each panel. Two boxes correspond to the two regions of the footpoints of the hot channels in panels f of Figures \ref{0712}--\ref{0910}. The arrows in panels a and d denote the secondary flare ribbons.}
\label{footpoints}
\end{figure}

\begin{figure}
\center {\includegraphics[width=14cm]{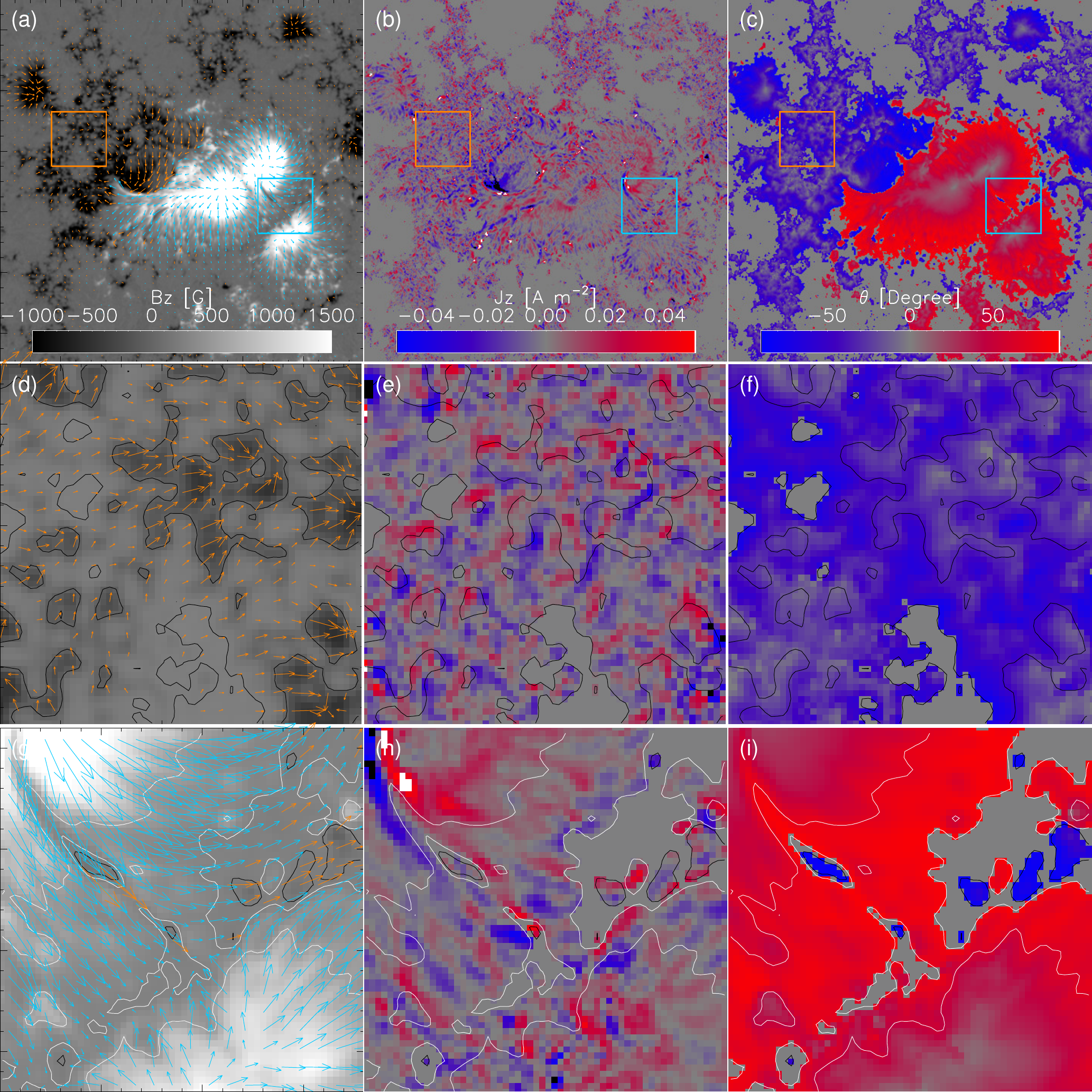}}
\caption{(a) HMI HARP cylindrical equal-area vector magnetegram of the AR 10720 at 15:58 UT on 2012 July 12. The background is the vertical magnetic field with the positive (negative) polarity plotted in white (black). The contours of the vertical magnetic field are also plotted. The arrows display the horizontal magnetic field. The box in blue (orange) indicates the positive (nagetive) footpoint of the hot channel. The size of the boxes (22 Mm$\times$22 Mm) is the same as that of the boxes (30\arcsec$\times$30\arcsec) in Figure \ref{0712}f. (b) Map of the vertical current density. (c) Map of the inclination angle of the magnetic field. The vertical current density and inclination angle at the areas with $B_z<$50 are set to be zero. (d)--(f) Vector magnetic field, vertical current density, and inclination angle at the negative footpoint of the hot channel. (g)--(i) Vector magnetic field, vertical current density, and inclination angle at the positive footpoint. The FOV of panels a--c is 145 Mm$\times$145 Mm.}
(Animation of this figure is available in the online journal.)

\label{0712_vector}
\end{figure}

In order to calculate the average magnetic field strength, average inclination angle, direct currents, and return currents at the footpoints, we choose two square regions with their centers determined from the ends of the hot channels but slightly shifted so as to cover most of the brightenings at the ends of the hot channels in the early rise phase as seen in the AIA 304 {\AA} and 1600 {\AA} passbands. The size of the square is set to be 30\arcsec$\times$30\arcsec (22 Mm$\times$22 Mm), which is big enough to contain all of the footpoints of the hot channels. On the other hand, such a size is also small enough for the sake of avoiding the influence of the surrounding magnetic field. Furthermore, we also compare the selected regions with the synthesized flare ribbons, which are derived through summing up all flare ribbons in the period of the early phase of the flares. We find that for each event, the two selected regions basically cover the hooks of the two J-shaped flare ribbons, in particular, the ribbons with stronger magnetic fields (Figure \ref{footpoints}). Finally, it is needed to determine the locations of the footpoints in the coordinate system of the HARPs data. Here, this is done by means of co-aligning the same magnetic features of the footpoints in the light-of-sight magentograms with that in the vertical component of the HARPs vector magnetograms. It is worth noticing that the real size (30\arcsec$\times$30\arcsec) of the two square regions in the light-of-sight magnetograms is slightly larger than the size (22 Mm$\times$22 Mm) of the corresponding two square regions in the HARPs vector magnetograms because the former has a projection effect. However, we vary the center and/or size of the region by within $\pm$5 Mm as a test and do not find that our results are influenced qualitatively although the values of magnetic parameters are changed slightly. Note that, for the SOL2012-07-12T and SOL2014-09-10 events, there appears a secondary ribbon \citep{zhangjun14_ribbon} that extends from the tip of the hook (arrows in Figure \ref{footpoints}). With the development of the flare, it further propagates toward a particular direction, forming a long and slender ribbon-like structure (see the attached online movies of Figure \ref{0712} and \ref{0910}).

Table \ref{tb2} shows the magnetic properties at the two regions where the hot channels root in. We find that at the footpoints with stronger magnetic fields, the average inclination angle tends to be larger with the maximal (minimal) value being 45$^\circ$ (23$^\circ$). The direct current there is calculated to be of the order of 10$^{12}$ A. While at the footpoints with weaker magnetic fields, the average inclination angle tends to be smaller with the maximal (minimal) value being 34$^\circ$ (18$^\circ$). The direct and return current is distributed nearly uniformly without obvious concentrations in some specific regions. Only for the SOL2012-07-12T event, the direct current there is comparable in quantity to that at the footpoints with stronger magnetic fields. One may expect that, the direct current flowing into the MFRs from one footpoint should flow out from the other, or that the direct current at the footpoints with weaker magnetic fields should be roughly equal to that at the footpoints with stronger magnetic fields. However, the real situation is complicated. Some potential reasons that cause the direct current at the footpoints with weaker magnetic fields significantly deviating from that at the other ones probably include: (1) the measurement of the vector field is not precise enough in the weak field region, which leads to a relatively large uncertainty in the current calculation there, and (2) the magnetic field strength is different at two different polarities, which requires that the sizes of regions for integrating the direct currents are also different to ensure the magnetic flux balance. However, observationally, the real sizes of the regions that contain the entire footpoints of MFRs can hardly be determined accurately. 

\begin{figure}
\center {\includegraphics[width=16cm]{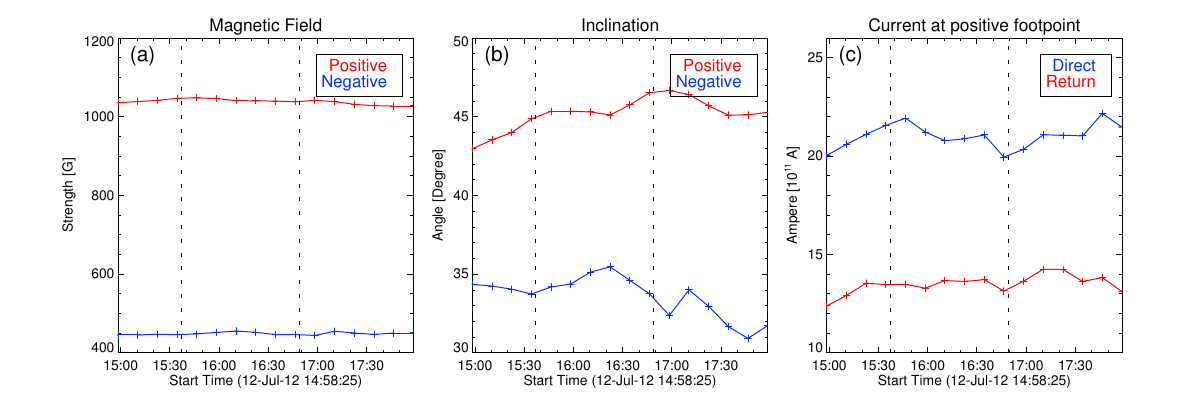}}
\caption{(a) and (b) Temporal evolution of the average magnetic field strength and inclination angle at the positive (red) and negative (blue) footpoints. (c) Temporal evolution of the absolute value of the direct (blue) and return (red) current at the footpoint of the hot channel with stronger magnetic field. Two vertical lines denote the onset and peak times of the flare.}
\label{0712_para}
\end{figure}

\begin{figure}
\center {\includegraphics[width=14cm]{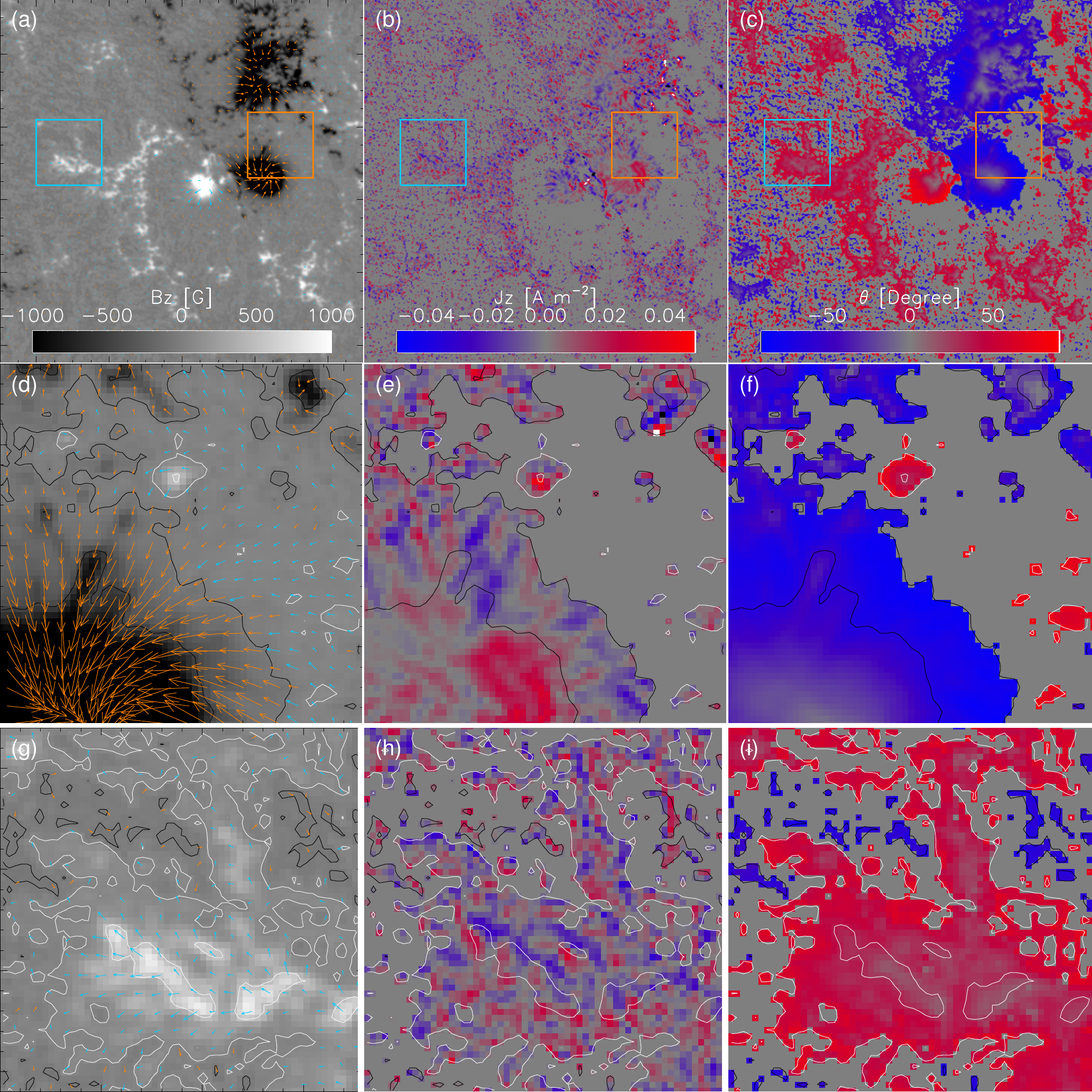}}
\caption{Same as Figure \ref{0712_vector} but for AR 11719 at 06:58 UT on 2013 April 11. The FOV of panels a--c is 120 Mm$\times$120 Mm. The FOV of panels d--i is 22 Mm$\times$22 Mm.}
(Animation of this figure is available in the online journal.)

\label{0411_vector}
\end{figure}

\begin{figure}
\center {\includegraphics[width=16cm]{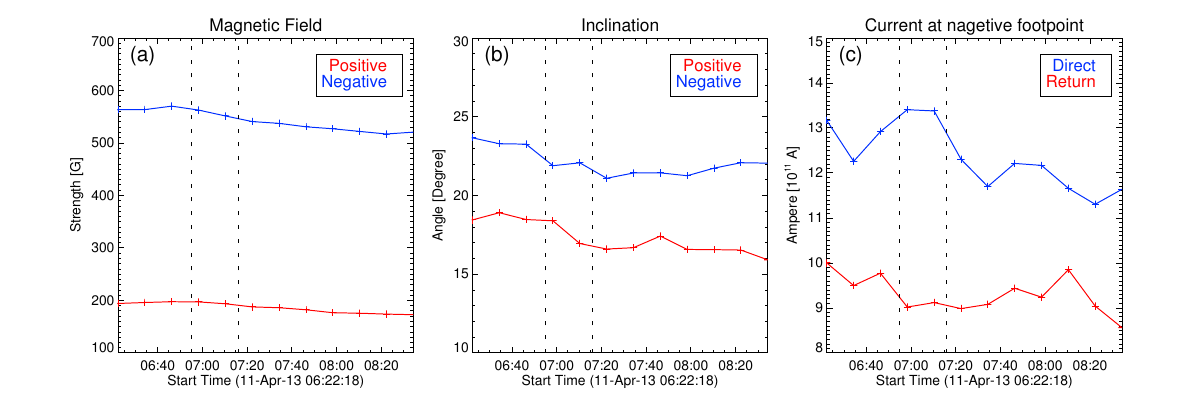}}
\caption{Same as Figure \ref{0712_para} but for the SOL2013-04-11T event.}
\label{0411_para}
\end{figure}

\begin{figure}
\center {\includegraphics[width=14cm]{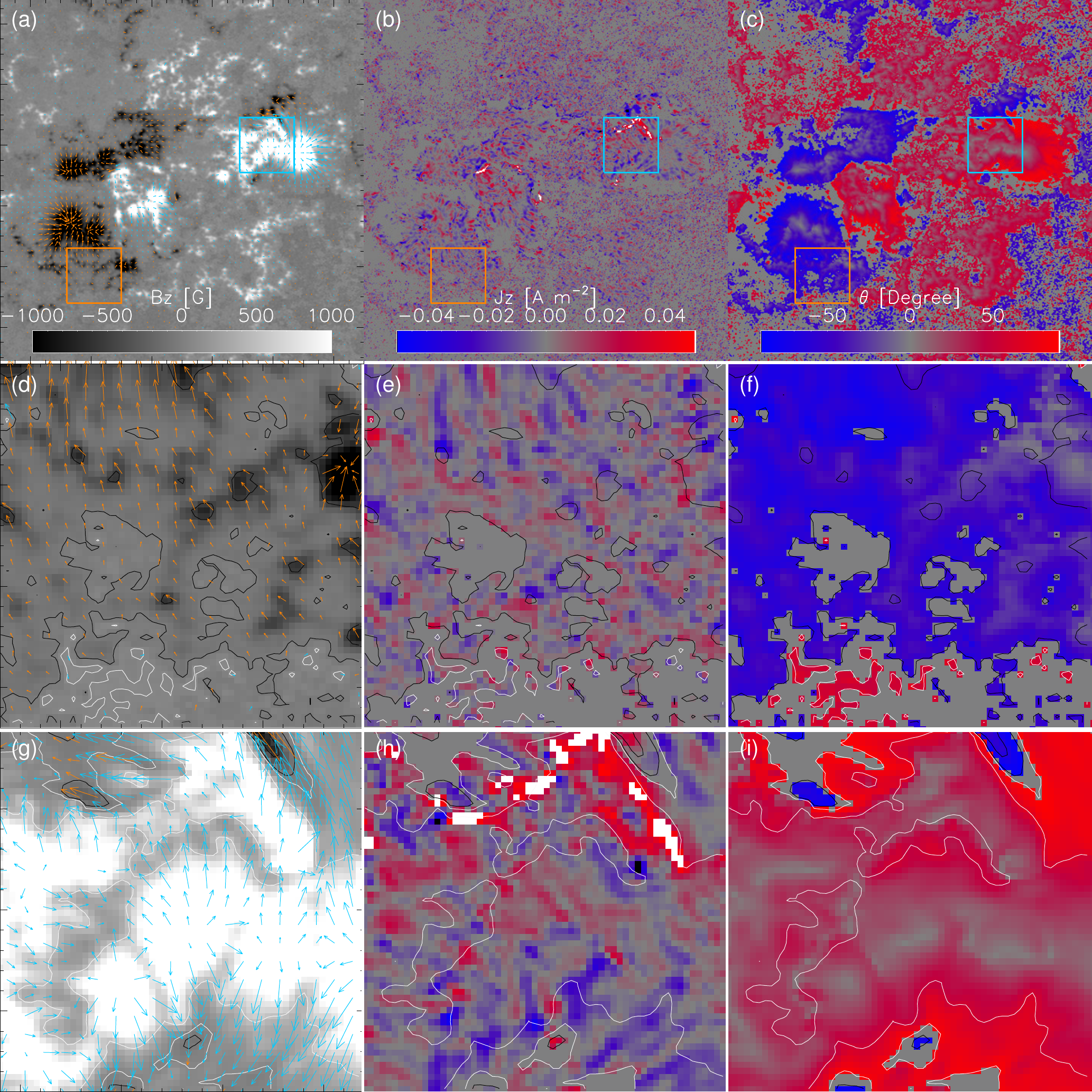}}
\caption{Same as Figure \ref{0712_vector} but for AR 12036 at 12:34 UT on 2014 April 18.}
(Animation of this figure is available in the online journal.)

\label{0418_vector}
\end{figure}

\begin{figure}
\center {\includegraphics[width=16cm]{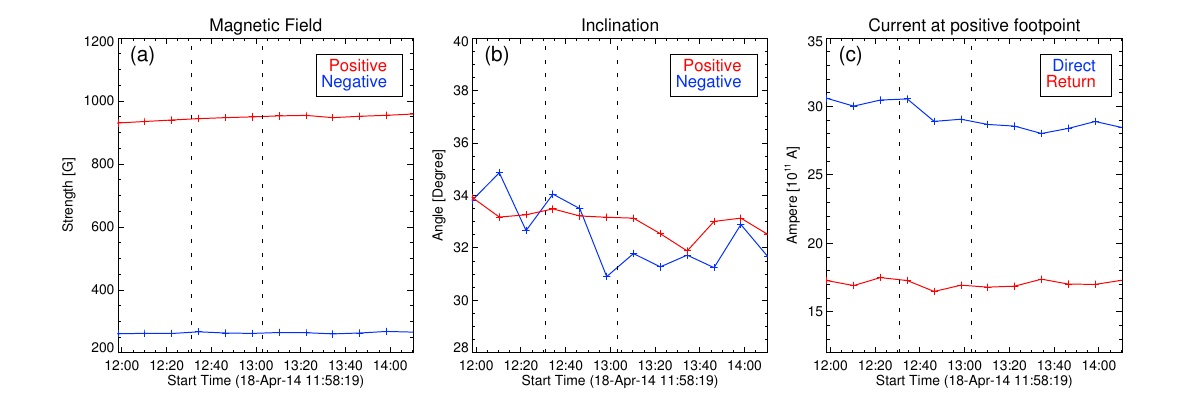}}
\caption{Same as Figure \ref{0712_para} but for the SOL2014-04-18T event.}
\label{0418_para}
\end{figure}

\begin{figure}
\center {\includegraphics[width=14cm]{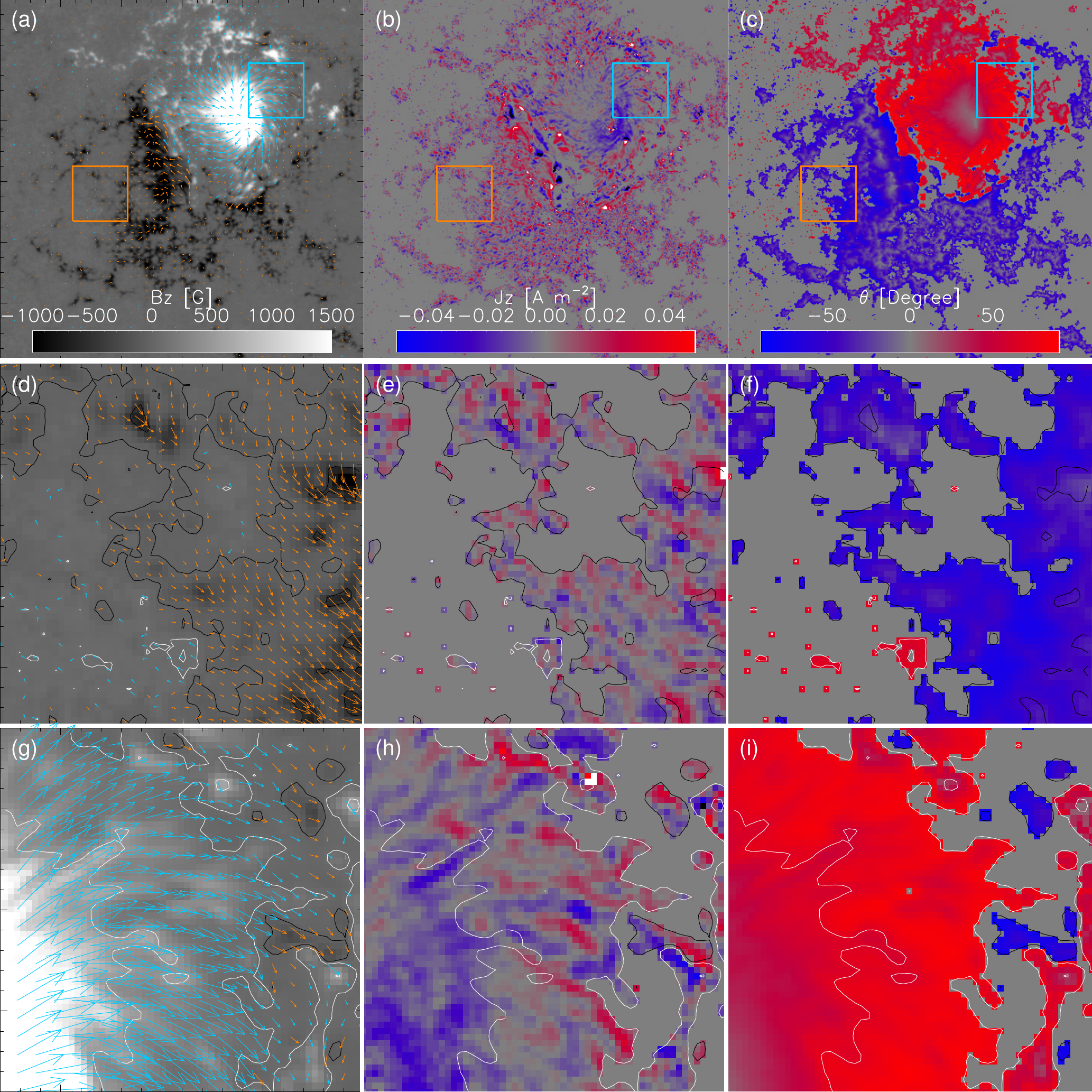}}
\caption{Same as Figure \ref{0712_vector} but for AR 12158 at 16:58 UT on 2014 September 10.}
(Animation of this figure is available in the online journal.)

\label{0910_vector}
\end{figure}

\begin{figure}
\center {\includegraphics[width=16cm]{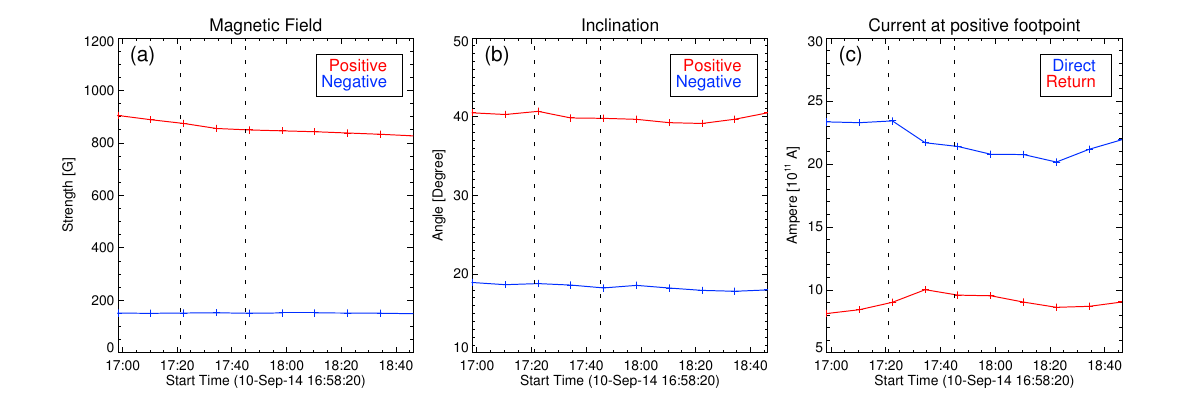}}
\caption{Same as Figure \ref{0712_para} but for the SOL2014-09-10T event.}
\label{0910_para}
\end{figure}

We also study the evolution of the average magnetic field strength, average inclination angle, and direct current at the two footpoints during the eruptions of the hot channels (Figures \ref{0712_para}, \ref{0411_para}, \ref{0418_para}, and \ref{0910_para}). The average magnetic field strength has no significant change with the eruptions, as a consequence of no obvious flux emergence or magnetic cancellation appearing. However, we find that the average inclination angle at most of the footpoints tends to decrease. For example, at the two footpoints of the SOL2013-04-11T and SOL2014-04-18T hot channels, the inclination angle significantly decreases. While the inclination angle only slightly decreases at the footpoints of the SOL2014-09-10T hot channel and it even increases at the footpoint of the SOL2012-07-12T hot channel with positive magnetic polarity. We suspect that such an increase is the result of the rapid rotation of the preceding sunspot \citep[see Figure 7 of ][]{cheng14_formation}, which compensates for the decrease of the inclination angle caused by the eruption of the hot channel. Moreover, we also find that the direct current at all of the footpoints with stronger magnetic fields experiences an obvious decrease (Figures \ref{0712_para}c, \ref{0411_para}c, \ref{0418_para}c, and \ref{0910_para}c). These features indicate that the magnetic field emanating from the footpoints tends to become more straight and less twisted after the eruptions. This is reasonable because the legs of the hot channel-like MFRs are gradually stretched and straightened. At the same time, such a stretching causes an increase of the length of the hot channels. However, due to the conservation of the total twist, the twist per unit length just decreases, leading to the reduction of the direct current.

\subsection{Evolution of Footpoints of the MFR Eruptions}
The detailed kinematical studies suggest that the MFRs play an important role in forming the CMEs and producing the flare emission \citep{cheng13_driver,cheng13_double,sun15_nc}. However, during the MFR eruption, the detailed evolution of the magnetic field in the MFR and its relation to the ambient field are still unknown. In this Section, we pay our attention to the chromospheric brightenings caused by the four erupted hot channels, in particular studying their evolution and relation to flare ribbons.

\begin{figure}
\center {\includegraphics[width=11cm]{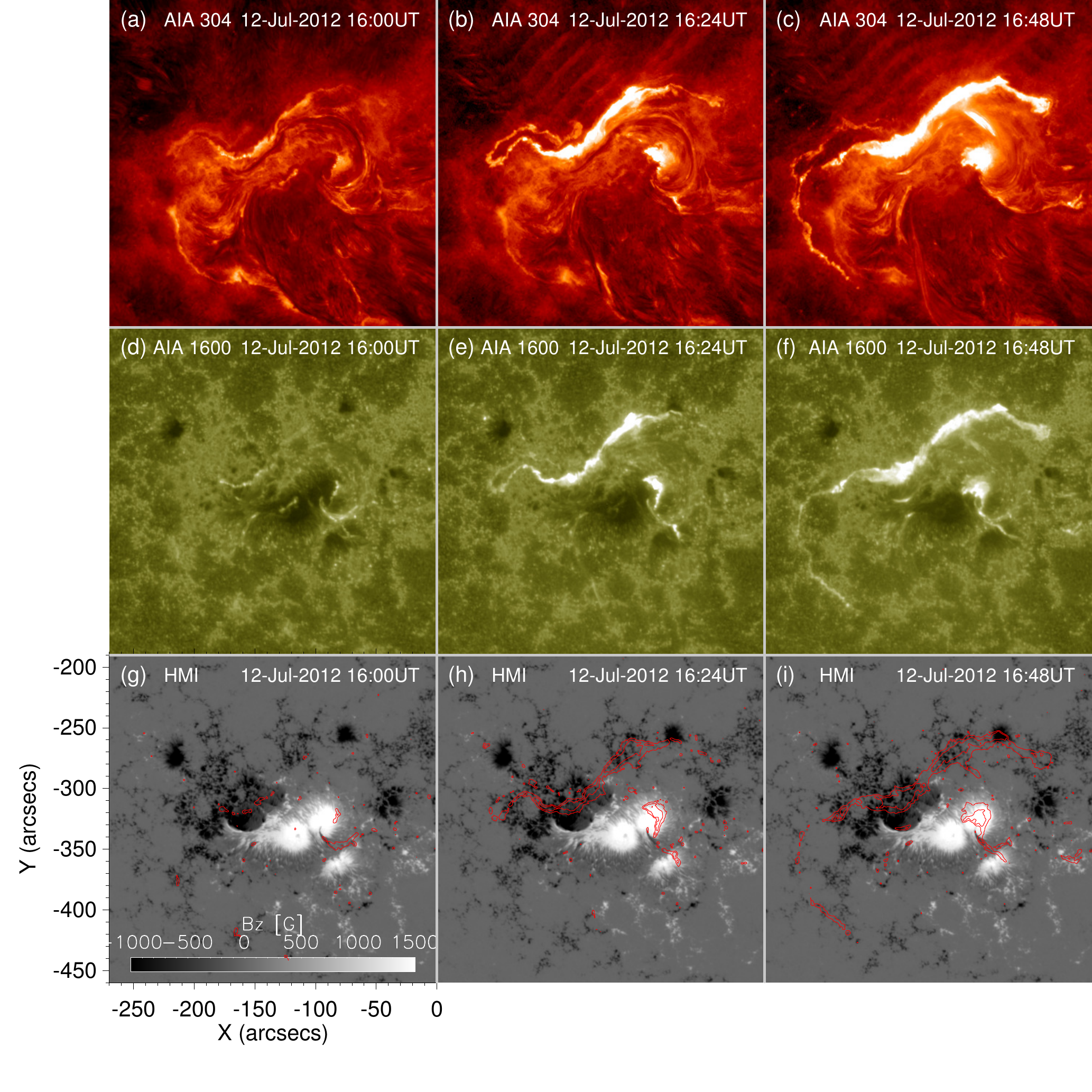}}
\caption{\textit{SDO}/AIA 304 {\AA} (a--c) and 1600 {\AA} (d--f) images showing the evolution of chromospheric brightenings in the SOL2012-07-12 event. The slender and long brightenings in the lower right corner of panels c and f indicate the secondary ribbon. The contours of the brightenings are also overlaied over the HMI light-of-sight magnetograms (g--i).}
\label{0712_1600}
\end{figure}

\begin{figure}
\center {\includegraphics[width=11cm]{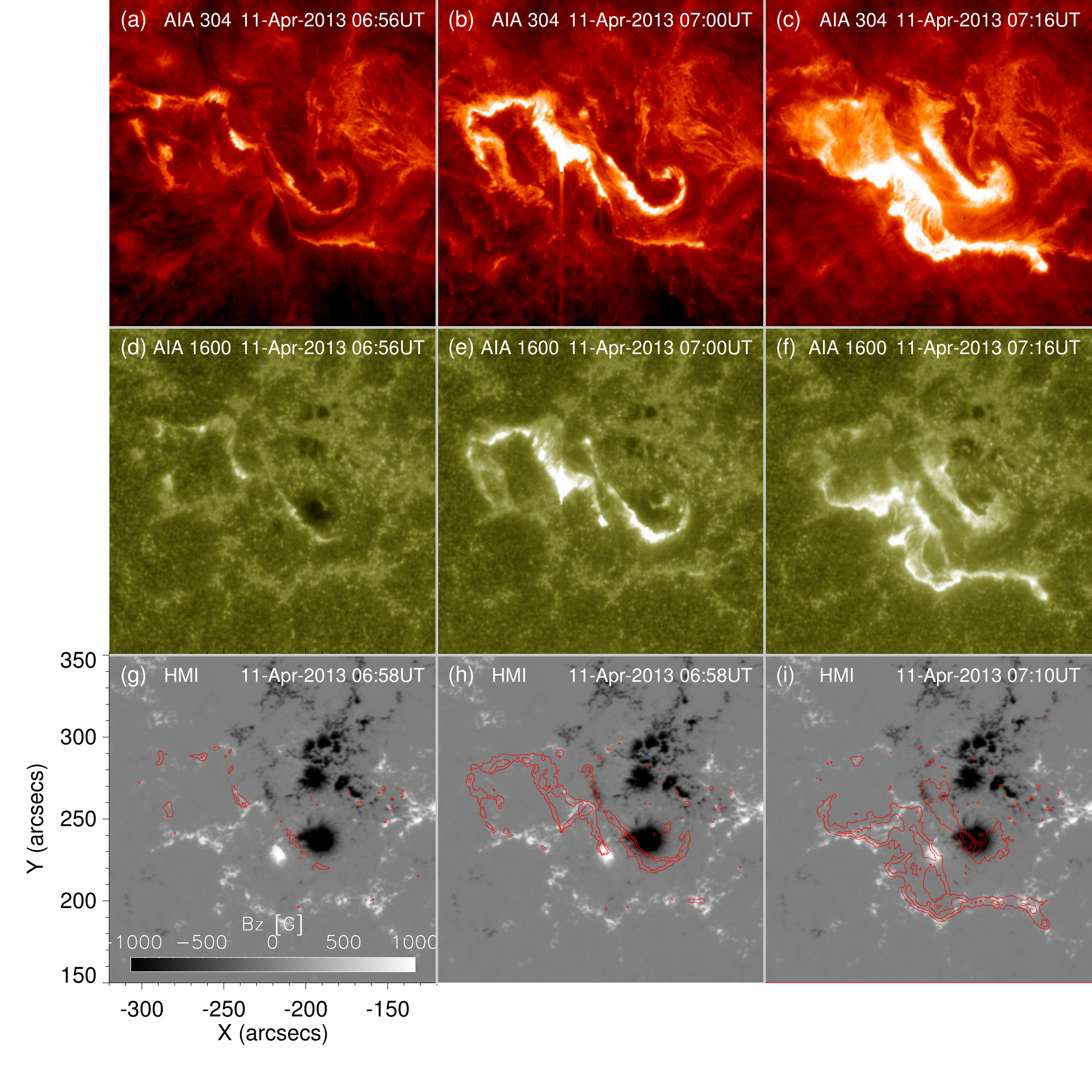}}
\caption{Same as Figure \ref{0712_1600} but for the SOL2013-04-11T event.}
\label{0411_1600}
\end{figure}

\begin{figure}
\center {\includegraphics[width=11cm]{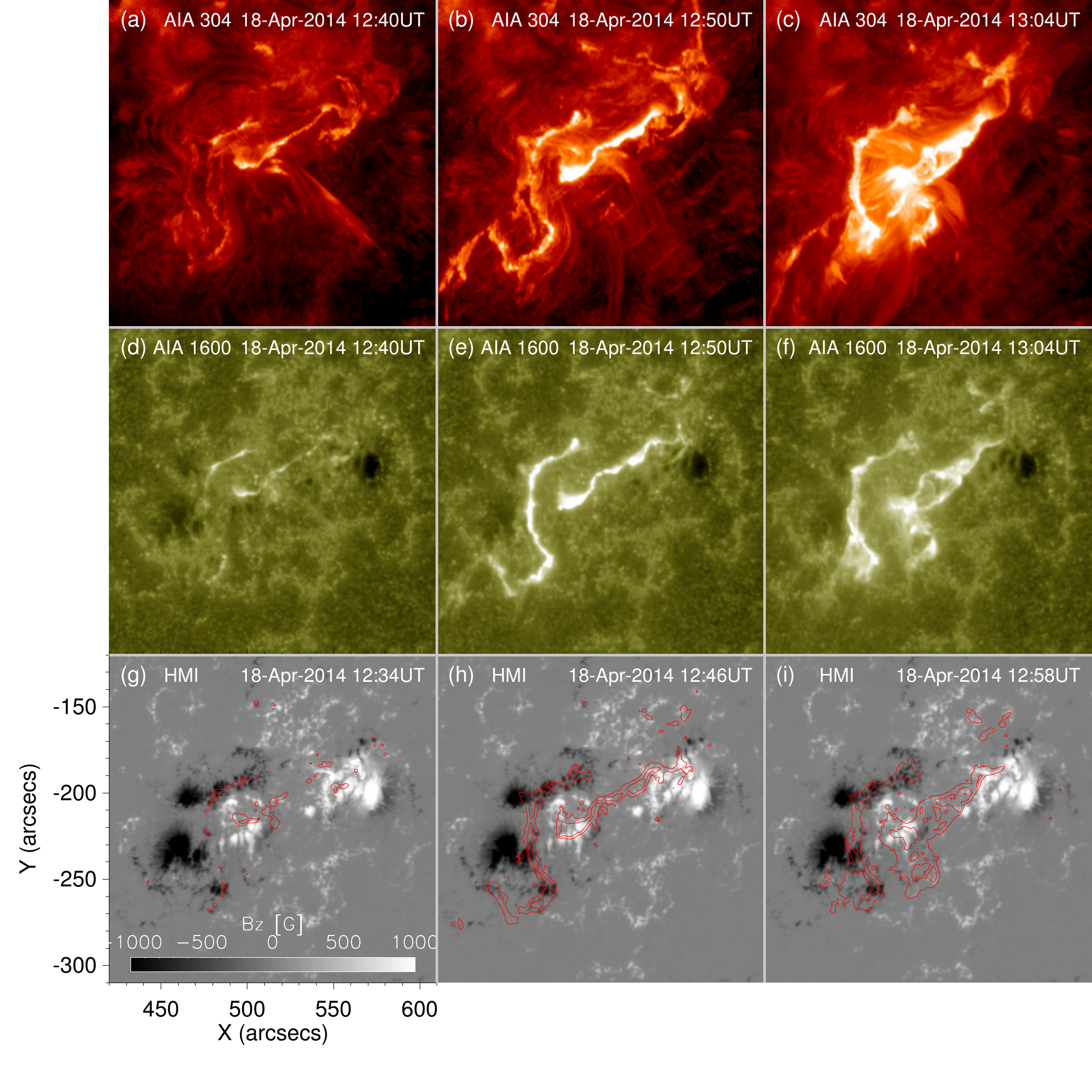}}
\caption{Same as Figure \ref{0712_1600} but for the SOL2014-04-18T event.}
\label{0418_1600}
\end{figure}

\begin{figure}
\center {\includegraphics[width=11cm]{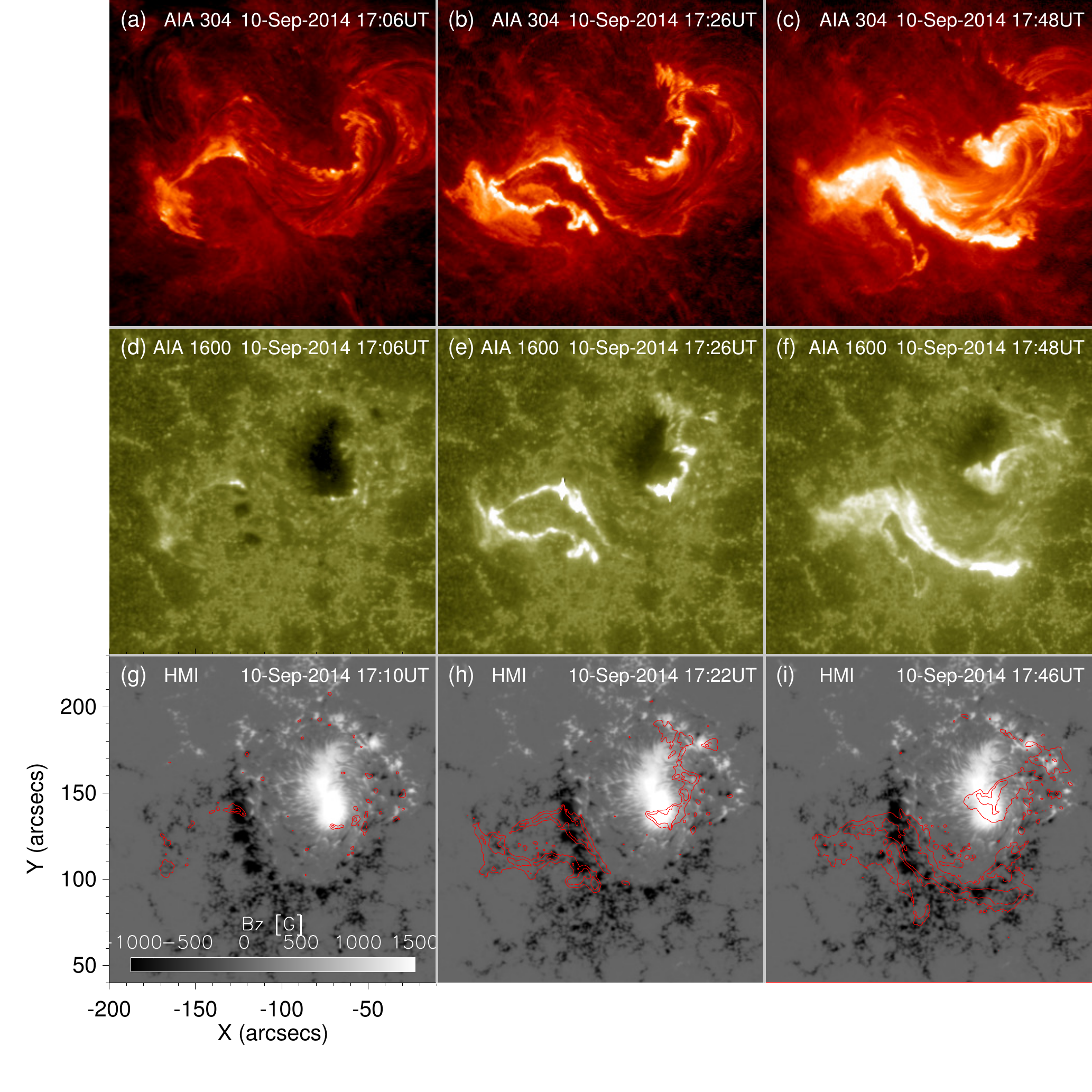}}
\caption{Same as Figure \ref{0712_1600} but for the SOL2014-09-10T event.}
\label{0910_1600}
\end{figure}

During the eruptions, the overall evolution of the chromospheric brightenings can be divided into three stages. The first stage occurs in the slow rise phase. The brightenings sporadically appear at the footpoints and in the regions below the two elbows of the hot channels (panels a, d, and g of Figure \ref{0712_1600}--\ref{0910_1600}). It indicates that some pre-heating probably has started, most likely taking place within the threads of the hot channels or between the hot channels and the ambient sheared arcades. 

The second stage of chromospheric brightenings starts with the beginning of the impulsive acceleration of the hot channels, also the early rise phase of the flare (panels b, e, and h of Figure \ref{0712_1600}--\ref{0910_1600}). The brightenings at the two footpoints first extend outward, forming a hook-like shape. At the same time, the brightenings also propagate along the main PILs with opposite orientations at two different polarities. Quickly, two slender ribbon-like brightenings come into being, taking on sheared double-J shape, the straight parts of which are adjacent to each other on either side of the PILs and the two hooks face on each other at the opposite ends. Such characters are especially apparent for the SOL2013-04-11T and SOL2014-04-18T events. For the SOL2012-07-12T and SOL2014-09-10T events, the J ribbon in the negative polarity is much larger than that in the positive polarity, which is primarily due to that the magnetic field is very strong there and thus the brightenings only propagate with a relatively short distance. 
 
The third stage happens in the later rise and decay phase of the flare (panels c, f, and i of Figure \ref{0712_1600}--\ref{0910_1600}). The primary characteristics are the separation of the two slender ribbons and the gradual disappearance of the two hooks. From the attached online movies of Figures \ref{0712}--\ref{0910}, one can clearly see that the two straight parts of the J ribbons expand outward with opposite directions but almost perpendicular to the PILs. Based on the CSHKP model, such a separation is mainly caused by the systematic ascending of the reconnection site in the corona. Moreover, from the second to third stage, we find that the two straight parts of the double J structure evolve from greatly sheared to almost parallelized (less sheared). This is most likely the consequence of the flare reconnection that proceeds from the strongly sheared overlying field at a low site to a nearly potential one at a high site \citep{aulanier12,janvier13}. 

\section{Summary and Discussion}
In this paper, we investigate the magnetic properties of four erupted hot channel-like MFRs that initially appear as forward or reversed sigmoidal structures prior to the eruption as seen in the AIA 131 and 94 {\AA} passbands. Thanks to the high resolution of the AIA, the continuously sigmoidal threads that comprise the coherent and sigmoidal channels can be seen to detach from the source sigmoidal ARs and then erupt to become the CMEs. The ratio of the projected length of the hot channels to their footpoint separation distance ranges from 2.3 to 4.0, showing that the magnetic fields in the hot channels have a strong writhe.

The axis of the hot channels significantly deviates from the main PILs and associated filaments prior to the eruption. It shows that the channels have ascended to a high altitude in the corona and been likely separated from that of the filaments when approaching the eruptions. Using the optimization algorithm \citep{wheatland00,wiegelmann04}, we also extrapolate the 3D NLFFF structures in the four ARs. Unfortunately, we could not reproduce the field lines that fit spatially with the hot channels. The reasons could be the insufficient resolution of magnetic field measurement, in particular at the footpoints of the hot channels \citep{cheng14_formation}, or the preprocessing over-smoothes the vector field before doing the extrapolation. Moreover, the observed vector magnetograms often contain significant Lorentz and buoyancy forces, which actually do not satisfy the NLFFF assumption \citep{derosa09}, thus challenging the reliability of extrapolated structures. At present, the potential method that is likely able to reproduce the magnetic field resembling the hot channels is the MFR insertion method developed by \cite{vanBallegooijen04}, which inserts an MFR into the potential field and then perform a relaxation to fit the observations \citep[e.g.,][]{suyingna12,suyingna15,savcheva12a,savcheva12b,savcheva15}. In a follow up work, we plan to utilize this method to model the structures of the four ARs and then compare the results with observations. We will also try the promising Grad-Rubin method, which has been applied to the SOL2014-09-10 event by \citet{zhaojie16} and reproduced twisted field lines that are comparable with the sigmoidal threads of the hot channel.

Through comparing the AIA 131 {\AA} and 94 {\AA} images and HMI light-of-sight magnetograms, we find that one footpoint of the hot channels originates in the penumbra or penumbra edge while the other one is from the moss region. Such an observational fact can be compared with various numerical models. In the simulations of a pre-existing MFR emerging into the potential magnetic arcades \citep[e.g.,][]{fan04,fan10,leake13}, the two footpoints of the MFR are exactly located at the centers of two different polarities. The situation is somewhat different in the simulations by \cite{aulanier10} and \cite{amari10}, in which an initially potential and asymmetric bipolar field evolves by means of the shearing motions along the main PIL. The flux cancellation in a bald-patch and/or the reconnection in a quasi-separatrix layer (QSL) transform the sheared arcades into continuously sigmoidal field, finally forming an MFR. The footpoints of the MFRs are located at the outer boundaries of the two polarities while the axis of the MFRs is basically parallel to the main PIL. Recently, \cite{amari14} performed a cutting-edge data-driven simulation and completely reproduce that one footpoint of the MFR is located at the edge of the positive polarity and the other one stems from the moss region nearby the negative polarity. Unfortunately, for the event they modeled, there are no AIA observations; thus it is difficult to compare the magnetic field lines with the hot channels.  
 
We find that the net currents in SOL2012-07-12T and SOL2014-09-10T ARs significantly deviate from zero, but in the SOL2013-04-11T and SOL2014-04-18T ARs they almost vanish. Nevertheless, at the footpoints of the four hot channels, we always see a dominated direct current. It indicates that the direct currents in the MFRs are always larger than the return currents irrespective of whether the currents in their source ARs are neutralized or not. In the ARs with significant non-zero net currents, the MFRs may be directly originated from the net currents; while in ARs with almost neutralized net currents, some opposite currents possibly exist elsewhere and can counteract the currents of the MFRs when integrating the net currents over the entire polarity. In fact, in many CME/flare initiation models \citep[e.g.,][]{titov99,vanBallegooijen04,torok05,kliem06,savcheva12a}, the existence of the MFR-associated current is required; one can see a current that comes out from one footpoint of the MFR and returns to the other one. However, it is worth noting that the direct currents at the footpoints with weaker magnetic field are almost one order of magnitude smaller than that at the footpoints with stronger magnetic field. This is somewhat different from the result of \cite{janvier14_current}, in which both of the footpoints of the erupted MFR originate in the regions with very strong magnetic field and exhibit comparative direct currents. Such a distinction may be due to the fact that, in our cases, the measurement of the vector magnetic field at the footpoints where the magnetic field is weak seems not to be accurate enough, which may lead to a relatively large uncertainty in the current calculation.

We study the evolution of the magnetic field at the footpoints of the hot channels. The average magnetic field strength does not change apparently with the eruptions going on. However, the average inclination angle at most of the footpoints has a decrease. The direct current at all of the footpoints with stronger magnetic field also experiences a decrease. At first glance, these results seem to conflict with the previous results that the photospheric magnetic field near the PIL becomes more horizontal \citep{wanghm10,wanghm15,wangs12,liuchang12}, implying an increase of the inclination angle there, and the direct current tends to increase slightly after the eruptions \citep{janvier14_current}. We argue that the different results may refer to the different physical processes that occur at different locations. With the MFR erupting upward, the magnetic field rooted in the two ends should be stretched to be more and more vertical. Meanwhile, due to the total twist conservation, the twist per unit length decreases. This results in a decrease of both the inclination angle and the direct current. By contrast, the magnetic field near the PIL is first stretched upward and then reconnects. Afterwards, the reconnected field lines shrink toward the photosphere due to magnetic tension, thus leading to an increase in the inclination angle. However, as for the increase of the direct current near the main PIL, it is possibly a result of magnetic reconnection or due to some other unknown mechanisms.
 
We also investigate the chromospheric brightenings at the footpoints of the hot channel-like MFRs, as well as their evolution and relation to flare ribbons. The overall evolution of the chromospheric brightenings is comprised of three stages. The brightenings in the first stage appear in the slow rise phase and is located at the two footpoints, as well as in the regions below the two elbows of the hot channels. The brightenings at the second stage take on a double J shape with the two hooks at the opposite ends corresponding to the extended footpoints. Once entering the third stage, the two straight parts of double J-ribbons start to separate from each other, evolving from two sheared slender ribbons into parallel widened ones. 

\begin{figure}
\vspace{-0.20\textwidth}
\center {\includegraphics[width=16cm]{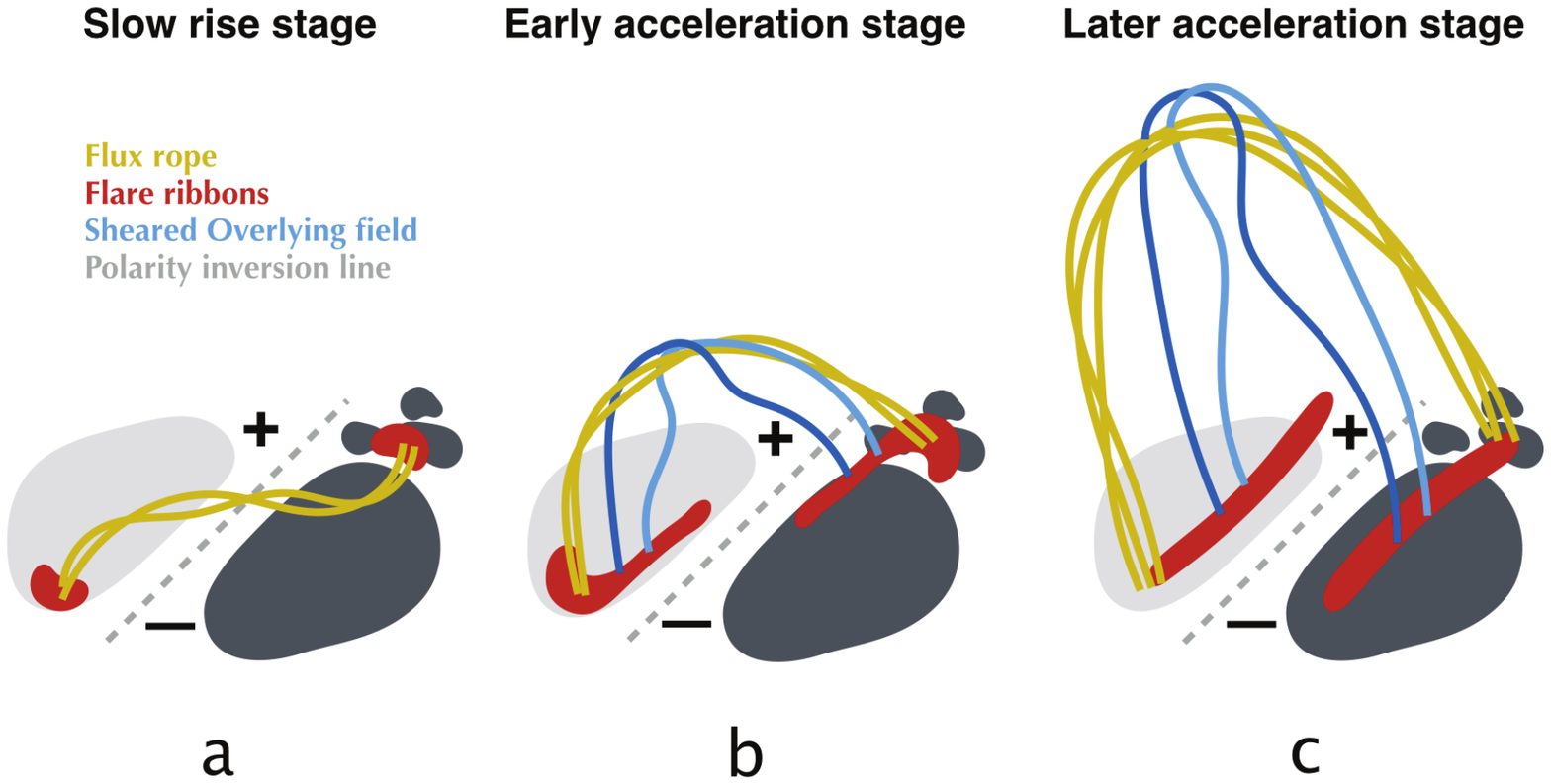}\vspace{-0.4\textwidth}}
\caption{Schematic drawing of the evolution of the magnetic field involved in the MFR eruption (yellow). The overlying field (blue) straddling over the MFR transits from being strongly to weakly sheared with the increasing height. The brightenings (red) in the chromosphere appear at the two footpoints of the MFR in the slow rise stage (a), then evolve into the sheared double J shape in the early acceleration stage (b), and finally form two parallel ribbons in the later acceleration stage (c). The gray (black) patches display the positive (negative) polarity of the AR with the small patches showing the moss region nearby.}
\label{cartoon}
\end{figure}

The evolution of the chromospheric brightenings provides a perspective on the 3D evolution of the MFR, which is illustrated by a schematic drawing as shown in Figure \ref{cartoon}. In the slow rise stage, the reconnection may take place inside the MFR or between the MFR and its ambient strongly sheared field. As a result, the footpoints of the MFR and the strongly sheared field are heated, manifesting themselves as some sporadical brightenings (Figure \ref{cartoon}a). At the same time, the reconnection also produces an upward Lorentz force to lift the MFR. Once the MFR ascends to a height where the decline of the background field is fast enough, kink and/or torus instability will be set in \citep[e.g.,][]{torok05,kliem06,olmedo10,savcheva12b,cheng13_double,zuccarello15,suyingna15}. Then, the MFR accelerates upward nonlinearly. In the early acceleration stage (Figure \ref{cartoon}b), the overlying field that straddles over the MFR is strongly sheared. When these fields are stretched by the MFR upward motion, their legs approach each other and then reconnect, resulting in two sheared J-shaped brightenings in the chromosphere. This result is consistent with previous findings that hard X-ray footpoint sources and post-flare loops sometimes display a strong shear in the initial phase of the flare \citep[e.g.,][]{suyingna07a,ji08,liuwei09,guo12}. Actually, the two strongly sheared J-shaped brightenings may represent the footprints of the curved QSLs in the chormosphere as expected by the 3D standard CME/flare model \citep{aulanier12,janvier13,janvier14_current,dudik16}. Very recently, through calculating the QSLs in several sigmoidal ARs, \cite{savcheva15} and \citet{zhaojie16} confirmed that the straight segments of the two sheared J-ribbons are basically matched by the MFR-related QSLs, thus approving the interpretation of the reconnection between strongly sheared overlying field in the early phase of the eruption.

In the later acceleration stage (Figure \ref{cartoon}c), the most apparent evolution is the separation of two flare ribbons. Since the MFR has risen to a relatively high altitude, the overlying field is almost potential. When a row of such overlying field lines is stretched and then reconnects, the heated footpoints will form two parallel ribbon-like brightenings. With the MFR continuously rising, the overlying field being reconnected becomes higher and higher; correspondingly, the separation of their footpoints becomes larger and larger. As a consequence, the increase in the reconnection height is mapped as the separation of two flare ribbons.

\acknowledgements We are grateful to the referee for his/her constructive comments that significantly improved the manuscript. We also thank Bernhard Kliem, Jie Zhang, Tibor {T{\"o}r{\"o}k}, and Kai Yang for their helpful discussions. \textit{SDO} is a mission of NASAs Living With a Star Program. X.C. and M.D.D. are supported by NSFC under grants 11303016, 11373023, and NKBRSF under grant 2014CB744203.


\clearpage
\begin{table*}
\caption{Properties of flares and CMEs caused by four erupted hot channel-like MFRs.}
\label{tb1}{
\begin{tabular}{ccccccc}
\\ \tableline \tableline
 Events             & Onset  & Peak    & Magnitude & Location &Speed & Morphology  \\
                         & [UT]  & [UT]    &   &   & [km s$^{-1}$] &   \\
\hline
SOL2012-07-12T     & 15:37  & 16:49 & X1.4           & S13W03 & 885     & Halo  \\
SOL2013-04-11T     & 06:55  & 07:16 & M6.5           & N07E13 & 861     & Halo  \\
SOL2014-04-18T     & 12:31  & 13:03 & M7.3           & S20W34 & 1203     & Halo  \\
SOL2014-09-10T     & 17:21  & 17:45 & X1.6           & N11E05 & 1267     & Halo  \\
\tableline
\end{tabular}}
\end{table*}

\begin{table*}
\caption{Magnetic parameters of the footpoints of four hot channel-like MFRs.}
\label{tb2}{
\begin{tabular}{cccccc}
\\ \tableline \tableline
                   &Unite       &  SOL2012-07-12T        & SOL2013-04-11T    & SOL2014-04-18T        & SOL2014-09-10T  \\
\hline
$F_+$                   &[10$^{21}$ Mx]        &29.3       &4.0     &18.9   &15.7 \\
$F_-$                    &[10$^{21}$ Mx]        &--25.0     &--9.5  &--11.7 &--13.3\\
$R_{F}$                & ...                           &0.08       &0.41   &0.24   &0.08\\
sign[$\bold{B}$]                       & +/--                          &+            &--        &+        &+\\
\hline
Sigmoid            &...   & forward   &reversed   &forward   &reversed\\
$I_d$                   &[10$^{12}$ A]           &21.1       &11.0 &31.1   &--14.3\\
$I_r$                    &[10$^{12}$ A]           &--14.8    &--10.0    &--30.0 &10.6\\
$I_n$                   &[10$^{12}$ A]           &6.3         &1.0     &1.1      &--3.7\\ 
$R_I$                  & ...                             &0.24       &0.05   &0.02    &0.15\\
\hline
$l$                        &[Mm]                         &312        &200    &184     &183\\
$d$                       &[Mm]                         &78        &67    &73     &80\\
$R$                      &...                              &4.0          &3.0     &2.5      &2.3\\
\tableline 
$\bar{B}_-$         &[G]                      &--447         &--570      &--261        &--115\\
$f_{-}$         &[10$^{21}$ Mx]                      &1.8         &1.4      &0.8        &0.4\\
$I_{d-}$              &[10$^{12}$ A]     &--2.17      &1.29   &--1.23      &0.80\\
$I_{r-}$               &[10$^{12}$ A]     &1.69        &--0.98      &1.00      &--0.65\\
$I_{n-}$              &[10$^{12}$ A]     &--0.48      &0.31      &--0.23    &0.15\\
$\bar{\theta}_-$   &[Degree]            &--34         &--23       &--32       &--18\\
\tableline
$\bar{B}_+$         &[G]                     &1047        &198       &939       &888\\
$f_{+}$         &[10$^{21}$ Mx]                      &2.7         &0.5      &3.3        &1.9\\
$I_{d+}$              &[10$^{12}$ A]    &2.16         &--0.93      &3.05       &--2.33\\
$I_{r+}$               &[10$^{12}$ A]    &--1.35      &0.92    &--1.75     &0.84\\
$I_{n+}$              &[10$^{12}$ A]    &0.81        &0.01       &1.30       &--1.49\\
$\bar{\theta}_+$   &[Degree]           &45           &18          &33          &40\\
\tableline 
\end{tabular}}

\vspace{0.03\textwidth}
Note: $F_+$ ($F_-$) denotes magnetic flux in the positive (negative) polarity of ARs. $R_F$ refers to the imbalance of magnetic fluxes defined as ($F_+$+$F_-$)/($F_+$--$F_-$). sign[$\bold{B}$] indicates the sign of the magnetic polarity where we calculate direct currents $I_d$, return current $I_r$, and net current $I_n$=$I_d$+$I_r$. $R_I$ is the ratio of the net current to total unsigned current and defined as $I_n$/($I_d$--$I_r$). The quantities $l$ and $d$ are the projected length and footpoint separation of the hot channels, respectively, and $R$=$l$/$d$. $\bar{B}_-$ ($\bar{B}_+$), $f_{-}$ ($f_{+}$ ), $I_{d-}$ ($I_{d+}$), $I_{r-}$ ($I_{r+}$), $I_{n-}$ ($I_{n+}$), and $\bar{\theta}_-$ ($\bar{\theta}_+$) represent average magnetic field strength, magnetic flux, direct current, return current, net current, and average inclination angle at the negative (positive) footpoints of the four hot channels, respectively.
\end{table*}

\clearpage

\clearpage

\end{document}